\documentclass[twocolumn]{aastex701}

 \usepackage{amsmath}	
 \usepackage{amssymb}	
 \usepackage{footnote}
 \usepackage{xspace}
 \usepackage{natbib}

\shorttitle{Winding-flux-based Masked PILs in Extreme Solar Flares}
\shortauthors{Halder et al.}
\graphicspath{{./}{figures/}}

\begin{document}

\title{Magnetic Topological Evolution and the Predictive Role of Winding-flux-based Masked PILs in Extreme Solar Flares from AR13664 and AR13842}

\correspondingauthor{Sudheer K. Mishra, Prithish Halder}
\email{sudheer1018@gmail.com, prithishh3@gmail.com, phalder2@unl.edu} 

\author[0000-0002-1073-1419]{Prithish Halder}
\affiliation{Department of Physics and Astronomy, Jorgensen Hall, University of Nebraska, Lincoln, NE 68588, USA}
\email{prithishh3@gmail.com, phalder2@unl.edu (Prithish Halder)} 

\author[0000-0003-2129-5728]{Sudheer K. Mishra}
\affiliation{Astronomical Observatory, Kyoto University, Sakyo, Kyoto 606-8502, Japan}
\email{sudheer1018@gmail.com} 

\author[0000-0002-1641-1539]{Abhishekh K. Srivastava}
\affiliation{Department of Physics, Indian Institute of Technology (BHU), Varanasi-221005, India}
\email{asrivastava.app@itbhu.ac.in} 

\author[0000-0002-5279-686X]{Ayumi Asai}
\affiliation{Astronomical Observatory, Kyoto University, Sakyo, Kyoto 606-8502, Japan}
\email{asai@kwasan.kyoto-u.ac.jp} 

\author[0000-0002-2198-4710]{Peter A. Dowben}
\affiliation{Department of Physics and Astronomy, Jorgensen Hall, University of Nebraska, Lincoln, NE 68588, USA}
\email{pdowben@unl.edu} 

\begin{abstract}
Using vector magnetogram data from the Helioseismic Magnetic Imager (HMI) on board the Solar Dynamics Observatory (SDO), we present detailed case studies of precursor activity in Active Regions (ARs) 13664 and 13842, which produced several intense flares. We examined the spatiotemporal evolution of key magnetic topological parameters, including winding, helicity, winding-flux-based masked polarity inversion lines (PILs), and PIL gradients, from the precursor to the post-flare phases. The temporal evolution of the winding flux showed a significant enhancement during the precursor phase, and this trend was consistent across the seven X-class and five M-class flares ($\geq$M5.0). The spatial distribution of both raw and Gaussian-smoothed winding-flux enhancement is co-located with flare footpoint regions of highly twisted core fields, as indicated by the photospheric magnetic topology. The winding-flux-based masked PILs that persisted around the flare locations showed a significant increase in strength and area during the precursor activity. These masked PILs spatially coincide with regions likely hosting non-potential core magnetic fields associated with the observed flares. Furthermore, the gradient of strong PILs extracted from the ARs was positively correlated with flare-associated locations and typically increased several hours (~3–5 h) before flare onset. By imposing a gradient threshold on strong PILs, we isolate magnetic regions repeatedly associated with flare initiation, providing useful constraints in regimes where current flare-forecasting models remain uncertain. Therefore, the combined analysis of the spatiotemporal evolution of different magnetic topological parameters enhances our understanding of intense solar flare forecasting and the identification of potential triggering sites, thereby contributing to improved space weather prediction frameworks. 

\end{abstract}


\section{Introduction} \label{sec:intro}

Solar flares are large-scale magnetic explosions in the solar atmosphere that release enormous amounts of stored magnetic energy (10$^{28}$–10$^{32}$ erg) as electromagnetic radiation, particle acceleration, and localized plasma heating \citep[see reviews][]{Benz2008, Benz2017, Shibata_et_al_2011, Webb_and_Howard_2012, Fletcher2024}. They often drive coronal mass ejections (CMEs), which in turn influence space weather conditions \citep{Schwenn2006, Gopalswamy2022}. Flares generally evolve in three phases: preflare, impulsive, and decay, depending on their energetic evolution. The preflare phase involves the accumulation of free magnetic energy driven by flux emergence, twisting, and shearing motions near the polarity inversion lines (PILs) \citep{Heyvaerts1977, SundaraRaman1998}. These processes increase magnetic complexity and non-potentiality, enabling the storage of substantial free energy in the field lines. This accumulated energy is subsequently released through magnetic reconnection or magnetohydrodynamic (MHD) instabilities \citep{Priest_and_Forbes_2002, Shibata_et_al_2011, 2018ApJ...856...86M, Srivastava_2019, Srivastava_2025, 2025ApJ...982..147M}.


Flux emergence and sunspot rotation drive the evolution of various magnetic topological parameters, such as magnetic helicity and relative helicity \citep{Park2008TheFlares, Yeats2018, Pariat2017RelativeEruptivity}, PIL properties \citep{schrijver_2007, Schrijver_2008, 2019ApJ...884..175W}, decay index and associated MHD instabilities \citep{Mitra2021}, as well as quasi-separatrix layers (QSLs) and regions prone to reconnection \citep{2012ApJ...750...15S}. It was found that active regions (ARs) with high helicity and magnetic energy are often associated with intense flares \citep{2012ApJ...759L...4T, 2014SoPh..289.4453M}. Examining the interplay between helicity injection, relative helicity, and the accumulation of free magnetic energy helps to understand the processes that may trigger solar flares and CMEs \citep{2010ApJ...718...43P, 2013ApJ...778...13P, 2012ApJ...759L...4T, 2019ApJ...887...64T}. In addition, these parameters serve as crucial diagnostics for improving flare forecasting \citep{Bobra_2015, 2017ApJ...835..156N, 2021JSWSC..11...39G, 2022ApJ...935...45S}. However, several studies have highlighted that helicity-based measures alone are not always reliable predictors, as both eruptive and non-eruptive ARs may exhibit similar high helicity and energy values \citep{Pariat2017RelativeEruptivity, Toriumi_2019}. This limitation has motivated the development of complementary topological parameters, such as magnetic winding, that can capture aspects of magnetic complexity not fully described by helicity.

The injection of helicity flux in ARs has been extensively investigated \citep{Demoulin2003MagneticLevel, Chae2001ObservationalFootpoints, LaBonte2007SurveyFlares, Liu2014MAGNETICREGIONS, Yamamoto2005MagneticLoops, Jeong2007MagneticRegions, Yang2009MagneticRegions, Chandra2010HowCloud, Park2012THEREGIONS, Romano2014EVOLUTIONROPES, Vemareddy2017Successive11928}, yet establishing a direct connection between helicity flux and emerging magnetic topology remains a challenge. This difficulty arises partly because magnetic helicity combines two aspects: field-line arrangement and magnetic flux. This fusion can complicate the interpretation of the helicity flux. For instance, a magnetic field with a simple, untwisted geometry but high intensity may yield a high helicity input without indicating the emergence of complex magnetic topology. In contrast, a strongly twisted topology with weak flux can produce a minimal helicity input, obscuring the actual configuration of field lines appearing in the atmosphere \citep{2003ApJ...593.1217P, Liu2012MAGNETICSUN, 2024A&A...686A.148S}.\\

To address the issue mentioned above, recent studies suggest that renormalizing the magnetic helicity flux by eliminating its dependence on magnetic flux magnitude can isolate the topological contribution of near-horizontal photospheric fields. It reveals the emergence of pretwisted flux and early signatures of flare- or CME-prone regions \citep{Prior2016TheFields, Prior2020MagneticFor, MacTaggart2021HelicityEmergence, MacTaggart2021DirectRegions, Aslam2024PhotosphericOnset}. This renormalization of the magnetic helicity, known as magnetic winding, is analogous to the localized helicity. Despite their conceptual similarity, the winding flux provides complementary information by capturing aspects of magnetic field topology that helicity alone cannot. Several theoretical and observational studies have further demonstrated the utility of winding flux in identifying preflaring activity \citep{2022ApJ...927..156R, MacTaggart2021DirectRegions, Alielden2023ARTop:Photosphere}. \citet{Mactaggart2024TheCME} studied the source region of the first halo CME that originated from the AR13664, demonstrating that magnetic winding flux is spatially coincident with the possible source region of the X1.0 and M7.2 class flares and significantly enhanced during the preflare phase. This finding shows the winding flux's capability to capture pre-flare characteristics and identify the possible source region of the flares. To fully harness its predictive power, it is essential to study the evolution of winding flux over time and establish its role in forecasting solar eruptive events. Time-series analysis of the winding flux and associated statistical parameters reveals that these topological parameters are useful for identifying pre-flaring activity in small- to large-scale flares \citep{Williams_2025}. These metrics can act as early indicators of magnetic energy buildup and impending flares.

While the winding flux provides a complementary measure of magnetic topology, its enhancements are often concentrated near PILs, which have long been recognized as prime sites for flare initiation. \citet{severny1958appearance} first established the significance of PILs in flare-productive ARs by analyzing the magnetogram data, showing that magnetic fields near PILs are typically highly sheared, twisted, non-potential, and characterized by steep gradients, making them prime candidates for large-scale flare triggering. Based on this insight, several studies have tracked the evolution of PILs during the preflaring phases of the flares \citep{hagyard_1990, falconer_2002, falconer_2006, falconer_2008, schrijver_2007, Wang_2008, Wang_2017, Toriumi_2013, Toriumi_2019, Toriumi_2024}. It is also important to note that converging and diverging motions near PILs contribute significantly to the storage of nonpotential (current-carrying) magnetic energy, which is eventually released in the form of small- to large-scale solar flares \citep{shimizu_2014, Toriumi_2017, sharykin_2020, Sun2021}. Due to these converging and divergent dynamics near the PILs, the weighted horizontal magnetic gradient (WHMG) method was introduced to forecast flares stronger than M5.0 \citep{2015ApJ...802L..21K}. The WHMG method enables the extraction of the geometric constraint for the location of flare emissions around PILs, ranging from small- to large-scale flares \citep{2019JSWSC...9A...6K, 2020ApJ...896..119K}. However, a comprehensive understanding of the spatiotemporal evolution of preflaring dynamics, including magnetic topology (e.g., helicity and winding flux), the formation and evolution of PILs associated with winding flux, and the development of strong PILs during the preflare phase, remains limited. Systematic investigation of how these features evolve and interact over time could provide critical insight into the physical conditions that govern the onset of intense solar flares.

In the present work, we investigate the preflare signatures of the two most flare-productive ARs of Solar Cycle 25: AR 13664 (May 2024) and AR 13842 (October 2024), which produced recurring X-class and several M-class flares. These highly productive ARs attracted several authors to investigate the source regions and infer the cause of such recurring flares \citep{Romano2024AnalyzingCharacteristics, Jarolim2024Magnetic13664, Jaswal2024Deconstructing10-11, Kontogiannis2024The13664, 2025ApJ...991...77M, 2025ApJ...985L..16D, 2025ApJ...985...20M}. In Section~2, we discuss the observational data and methodology, including the computation of the magnetic topology parameters. Section 3 presents the results, analyzing the temporal and spatial variations of different topological parameters and their correlation with flare onset time, possible source regions, and possible triggering sites. Finally, in Section 4, we discuss the broader implications of our findings, highlighting the potential of winding flux as a predictive tool for solar flares and its applications in improving space weather forecasting.

\section{Observational data and Analysis}

In this paper, we analyze AR13664 and AR13842, two highly flare-productive ARs during solar cycle 25, which first appeared on the solar disk on May 2, 2024, and September 28, 2024, respectively. Our study focused on intense flares (usually $\geq$ M5.0): five X-class and four M-class flares from AR13664 (May 8–11, 2024), and two major X-class and one M3.6-class flares from AR13842 (October 1–3, 2024). Both regions rapidly achieved complex $\beta\gamma\delta$-class configurations after their first appearance on the solar disk, showing intense and recurrent flare activity. Previous analyses of AR13664 demonstrate that successive bipolar flux emergence and shear motions near PILs drive free magnetic energy accumulation and flare initiation \citep{Jarolim2024Magnetic13664, 2024ApJ...973L..31R}. Detailed data-driven MHD simulations and observations discuss the cause of the triggering mechanism of these two X7.1 and X9.0-class flares in AR13842 \citep{2025ApJ...991...77M, 2025ApJ...985L..16D, 2025ApJ...985...20M}.
\subsection{Observational Data}

We utilized vector magnetogram data from the \textit{Helioseismic and Magnetic Imager (HMI)} \citep{2012SoPh..275..207S} onboard the \textit{Solar Dynamics Observatory (SDO)} \citep{2012SoPh..275....3P}, with 0.5\arcsec pixel size and 720~s temporal cadence \citep{Sun2013}. Our primary analysis used the Spaceweather HMI Active Region Patch (SHARP) dataset \citep{2014SoPh..289.3549B} for automated active-region detection and tracking. The SHARP dataset includes vector magnetograms remapped using a Lambert Cylindrical Equal-Area (CEA) projection, allowing the Cartesian field components [\textit{$B_x$, $B_y$, $B_z$}] to be transformed into the spherical heliographic coordinates [\textit{$B_{\phi}$, $B_{\theta}$, $B_r$}]. Such coordinate transformations are crucial for extracting helicity and magnetic winding. We also use extreme ultraviolet (EUV) observations from the \textit{Atmospheric Imaging Assembly (AIA)} \citep{2012SoPh..275...17L}, with a pixel size of 0.6\arcsec\ and a temporal cadence of 12~s to observe the coronal connectivity.

\begin{figure*}[ht]
    \centering
\includegraphics[trim = 0.0cm 0.0cm 0.0cm 0.0cm, scale=0.2]{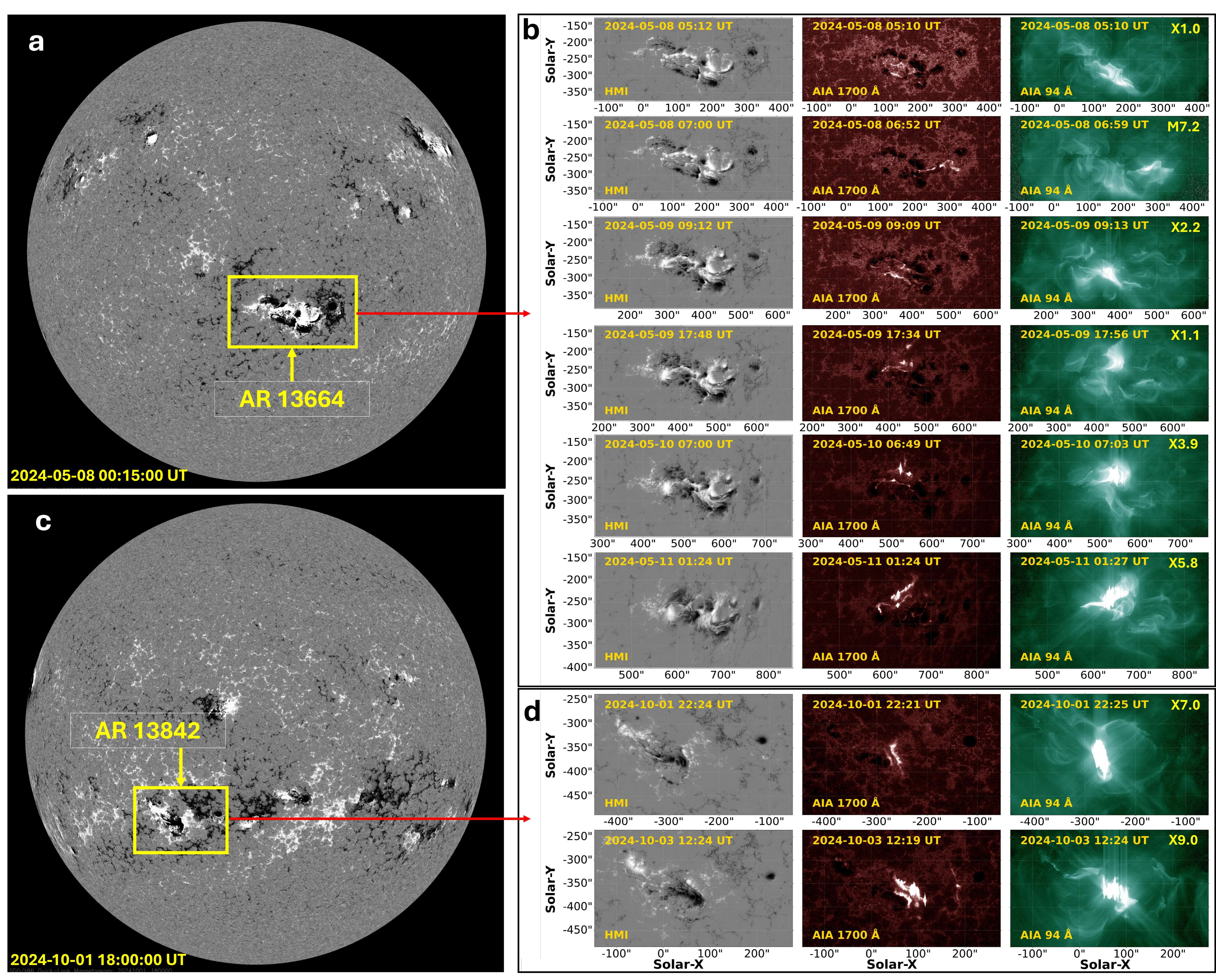}
   \caption{Panels (a) and (c) display the full HMI LOS magnetogram FOVs for NOAA AR13664 and AR13842, which produce recurring X-class flares. For each flare, panels (b) and (d) are zoomed-in views of HMI LOS magnetic field (left), AIA 1700 {\AA} flare ribbons (center), and AIA 94 {\AA} flare loops (right). Panel (b) highlights AR13664 during the X1.0 \& M7.2 flare on 2024 May 8; X2.2 \& X1.1 on May 9; X3.9 on May 10; and X5.8 on May 11. Panel (d) covers AR13842 during X7.1 and X9.0 flares on October 1 and 3, 2024.}
    \label{fig:overview}
\end{figure*}

\begin{figure*}
    \centering
\includegraphics[trim = 2.0cm 0.0cm 2.0cm 0.5cm, scale=0.7]{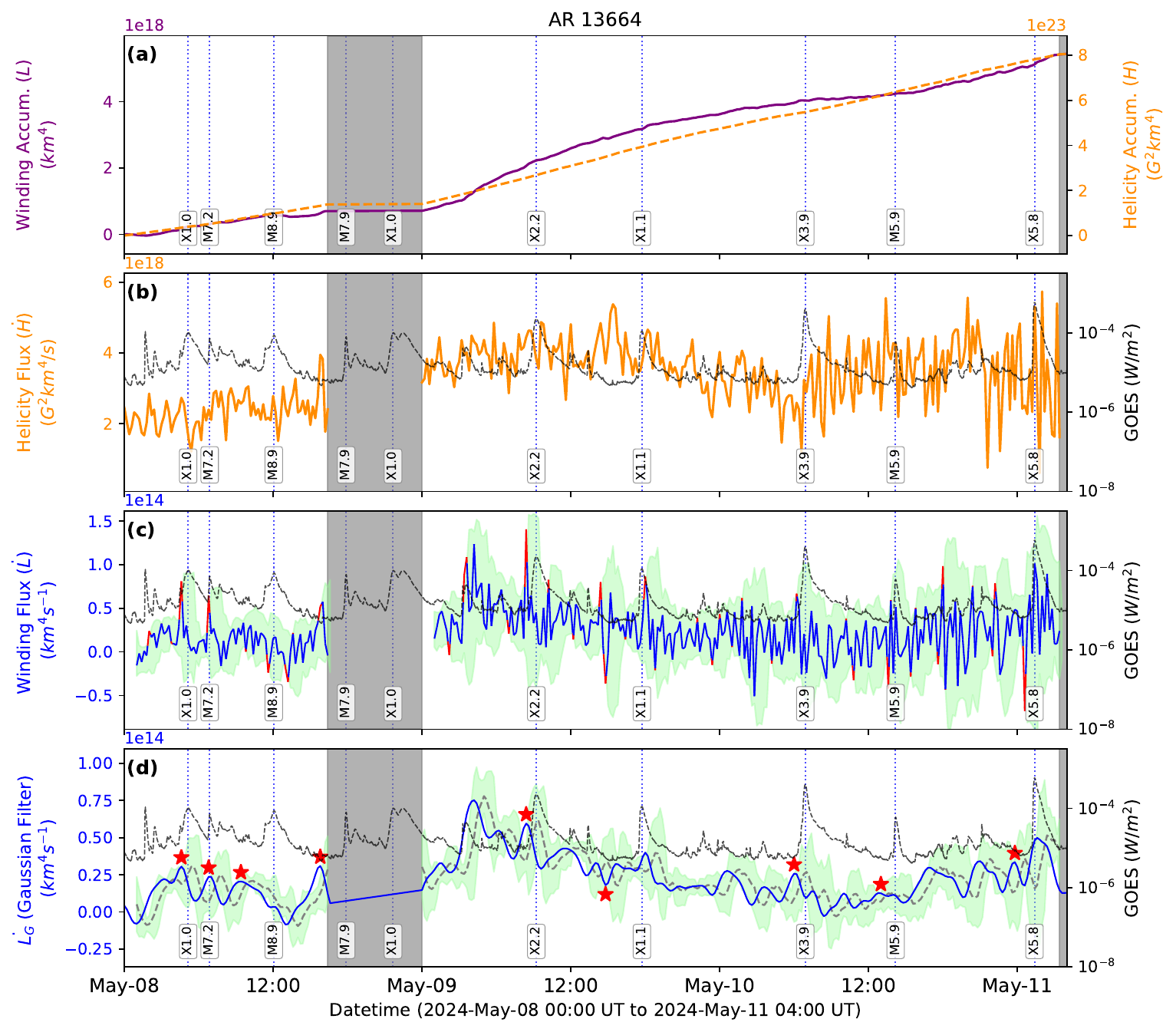}
\caption{Panel (a) shows the temporal evolution of helicity accumulation (dotted orange line) and winding accumulation (solid magenta line) in AR 13664 from 00:00 UT on May 8, 2024, to 04:00 UT on May 11, 2024. The GOES 1--8~\AA{} X-ray flux (dashed black line) is plotted against the right-hand $y$-axis in panels (b), (c), and (d). Vertical dotted lines indicate the peak times of flares $\ge$M5.0. Panel (b) displays the temporal evolution of the helicity flux ($\dot{H}$, orange line). Panel (c) shows the winding flux, $\dot{L}$ (green line; units of $\mathrm{km}^{4}\,\mathrm{s}^{-1}$), overlaid with the running mean, $\mu$, calculated from the preceding hour of data, and an envelope of $\mu \pm 2\sigma$ (green shaded region), where $\sigma$ represents the standard deviation. Significant excursions beyond the $2\sigma$ threshold are highlighted (see red lines in panel 'c'), indicating potential statistical precursor signatures. Panel (d) shows the temporal evolution of the winding flux smoothed using a three-dimensional Gaussian filter ($\dot{L}_{G}$; blue line) and backward-looking signal (the running mean of the preceding hour, $\mu$; dashed grey curve and a $\mu\pm2\sigma$ envelope; green shaded region), with red stars marking distinct excursions of the precursor indicator preceding the major flaring activity.}
    \label{fig:temporal_L_AR13664}
\end{figure*}

\begin{figure*}
    \centering
\includegraphics[trim = 0.0cm 0.0cm 0.0cm 0.0cm, scale=0.7]{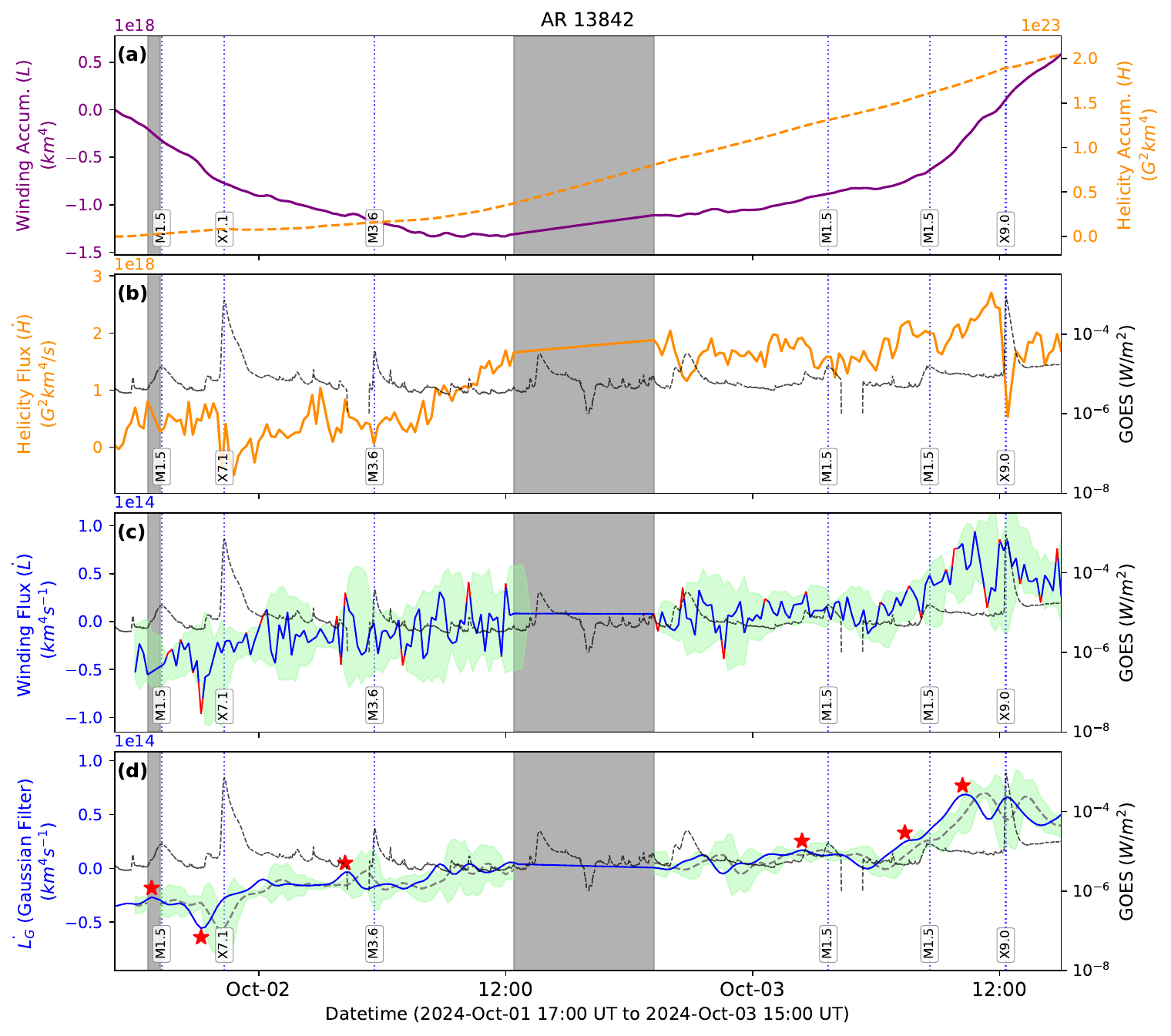}
\caption{Similar to Figure~2, but for AR~13842, covering the period from 2024 October 1 at 17:00~UT to 2024 October 3 at 15:00~UT.}
    \label{fig:temporal_L_AR13842}
\end{figure*}

\subsection{Magnetic Topology Calculations}
We use the open-source \texttt{ARTop (Active Region Topology)} software package\citep{Alielden2023ARTop:Photosphere}\setcounter{footnote}{0}\footnote{\texttt{ARTop:} \url{https://github.com/DavidMacT/ARTop}} to compute the magnetic helicity and winding fluxes from the SHARP vector magnetogram data. \texttt{ARTop} is an open-source tool designed to extract key topological parameters, such as magnetic helicity and winding, decompose them, and analyze time series, thereby enabling a detailed investigation of the spatiotemporal evolution of ARs. It uses the DAVE4VM technique \citep{Schuck2008TrackingEquation} to derive photospheric velocity fields from HMI SHARP magnetograms and evaluate topological quantities at the photospheric plane. Detailed theoretical background and complete mathematical formulation of the helicity and winding fluxes derived by \texttt{ARTop}, including their braiding and emergence components, are provided in \citet{2022ApJ...927..156R} and \citet{Alielden2023ARTop:Photosphere}. Here we use the standard definitions implemented in \texttt{ARTop} \citep{Alielden2023ARTop:Photosphere} to compute the total magnetic helicity (helicity accumulation, $H$) and helicity flux ($\dot{H}$) as well as the total magnetic winding (winding accumulation, $L$), winding flux ($\dot{L}$) and the field-line winding flux $\dot{\mathcal{L}}$.

 In \texttt{ARTop}, total helicity flux (\(\dot{H}\)) is defined as
\begin{equation}
    \frac{\mathrm{d}H}{\mathrm{d}t}
    = -\frac{1}{2\pi} \int_{P \times P} \frac{\mathrm{d}\Theta(\mathbf{x}, \mathbf{y})}{\mathrm{d}t}
    \, B_z(\mathbf{x})\, B_z(\mathbf{y}) \, \mathrm{d}^2\mathbf{x} \, \mathrm{d}^2\mathbf{y},
    \label{eq:helicity_flux}
\end{equation}
where \(B_z(\mathbf{x})\) and \(B_z(\mathbf{y})\) are the components of the magnetic field \(\mathbf{B}\) normal to the photospheric plane \(P\) (the \(X\)–\(Y\) plane). In this context, \(\mathbf{x}\) and \(\mathbf{y}\) are independent position vectors on the photospheric plane, and \(\mathrm{d}\Theta(\mathbf{x}, \mathbf{y})/\mathrm{d}t\) represents the relative angular velocity of the footpoint at \(\mathbf{y}\) with respect to the footpoint at \(\mathbf{x}\). This is not a rigid-body rotation about a single fixed central axis; rather, it measures the relative rotational motion (or mutual orbital motion) of any pair of magnetic flux elements in the plane. Even translational motion of one footpoint relative to another contributes to a change in their mutual relative angle \(\Theta(\mathbf{x}, \mathbf{y})\), mathematically encoding the topological linking and twisting of the coronal field lines connected to them. More details about the \texttt{ARTop} package and the above-mentioned parameters can be found in \citep{Alielden2023ARTop:Photosphere}.

Similarly, the magnetic winding flux (\(\dot{L}\)) is defined as follows:
\begin{equation}
    \frac{\mathrm{d}L}{\mathrm{d}t} = -\frac{1}{2\pi} \int_{P \times P} \frac{\mathrm{d}\Theta(\mathbf{x}, \mathbf{y})}{\mathrm{d}t} \sigma_z(\mathbf{x}) \sigma_z(\mathbf{y}) \, \mathrm{d}^2\mathbf{x} \, \mathrm{d}^2\mathbf{y},
    \label{eq:winding_flux}
\end{equation}
where \( \sigma_z(\mathbf{x}) \) is defined as:
\begin{equation}
\sigma_z(\mathbf{x}) =
\begin{cases}
1 & \text{if } B_z(\mathbf{x}) > 0 \text{ and } |\mathbf{B}(\mathbf{x})| > \varepsilon, \\
-1 & \text{if } B_z(\mathbf{x}) < 0 \text{ and } |\mathbf{B}(\mathbf{x})| > \varepsilon, \\
0 & \text{if } B_z(\mathbf{x}) = 0 \text{ or } |\mathbf{B}(\mathbf{x})| \leq \varepsilon.
\end{cases}
\label{eq:polarities}
\end{equation}
where \(\varepsilon\) is the cut-off value of the field strength. We use \(\varepsilon\) = $\pm$50 G as the cut-off value of the field strength for both ARs in this paper.

We performed a spatio-temporal analysis of topological parameters using custom \texttt{Python} code to identify preflaring activity via winding flux, winding-flux-based masked PILs, and PIL gradients, thereby isolating flare-triggering sites. Details of the winding flux-based masked PIL extraction are discussed in Subsection~3.4. 

\section{Observational Results}
\subsection{Multiwavelength View of the Flares}
Panels (a) and (c) of Figure \ref{fig:overview} show the full-disk HMI line-of-sight (LOS) magnetograms, highlighting AR 13664 and AR 13842 within the yellow boxes. Panel (b) of Figure \ref{fig:overview} presents three vertical columns that provide a zoomed-in view of the complex magnetic structures and the peak flaring activity in AR 13664, using observations from the HMI-LOS magnetogram, AIA 1700 and AIA 94 {\AA}. The detailed timelines of the five X-class and four M-class flares analyzed from AR 13664 are provided in Table~1. Similarly, panel (d) presents a zoomed-in view of the source region of AR 13842, extracted from the yellow box of panel (c), which produced X7.1 and X9.0-class flares on October 1 and 3, 2024. In AR13842, we focus on two X-class and one M3.6-class flares (see Table~1 for a detailed timeline). We derive the magnetic field and velocity pattern components from vector magnetogram data to deduce the topological evolution of active regions and associated topological parameters using \texttt{ARTop}, and discuss the observational results in the following subsections.

\subsection{Temporal Variation of Winding Flux} \label{sec:temporal}
Figures \ref{fig:temporal_L_AR13664} and \ref{fig:temporal_L_AR13842} depict the temporal evolution of the magnetic helicity and winding parameters for AR13664 and AR13842, respectively. Panel (a) shows the long-term buildup of winding accumulation ($L$) and helicity accumulation ($H$), while panel (b) presents the helicity flux ($\dot{H}$). Both $H$ and $L$ are time-integrated quantities that trace the continuous, net accumulation of magnetic stress and topological complexity injected into the active region. As shown in panel (a) of Figure~\ref{fig:temporal_L_AR13664}, both parameters exhibit an overall gradual increase throughout the observing window, reaching their maximum values toward the end of the time domain. They do not exhibit discrete local pre-flare maxima or abrupt step-like drops immediately prior to individual X-class flares. This absence of isolated pre-flare peaks in the cumulative curves is largely due to the nature of the parameters and our selected observational baseline. We specifically restricted our study of AR\,13664 to May 8--11, 2024, when the region was near the disc centre to minimise projection and limb effects when estimating parameters from the SHARP dataset. During this window, AR\,13664 was continually evolving and intensely flare-productive, initiating major flare activity within the first 5 hours and producing recurrent X- and M-class flares throughout and beyond. Because the active region was undergoing continuous, successive stress injection without quiet relaxation intervals, the cumulative quantities continuously integrate the net stress rather than resetting or plateauing between flares. Thus, the accumulated measures reflect the macroscopic prerequisite budget of stored non-potentiality required to sustain recurrent high-energy activity, rather than serving as sensitive indicators for the precise timing or onset of individual eruptions.

\begin{figure*}
    \centering
\includegraphics[trim = 0.0cm 0.0cm 0.0cm 0.0cm, scale=0.27]{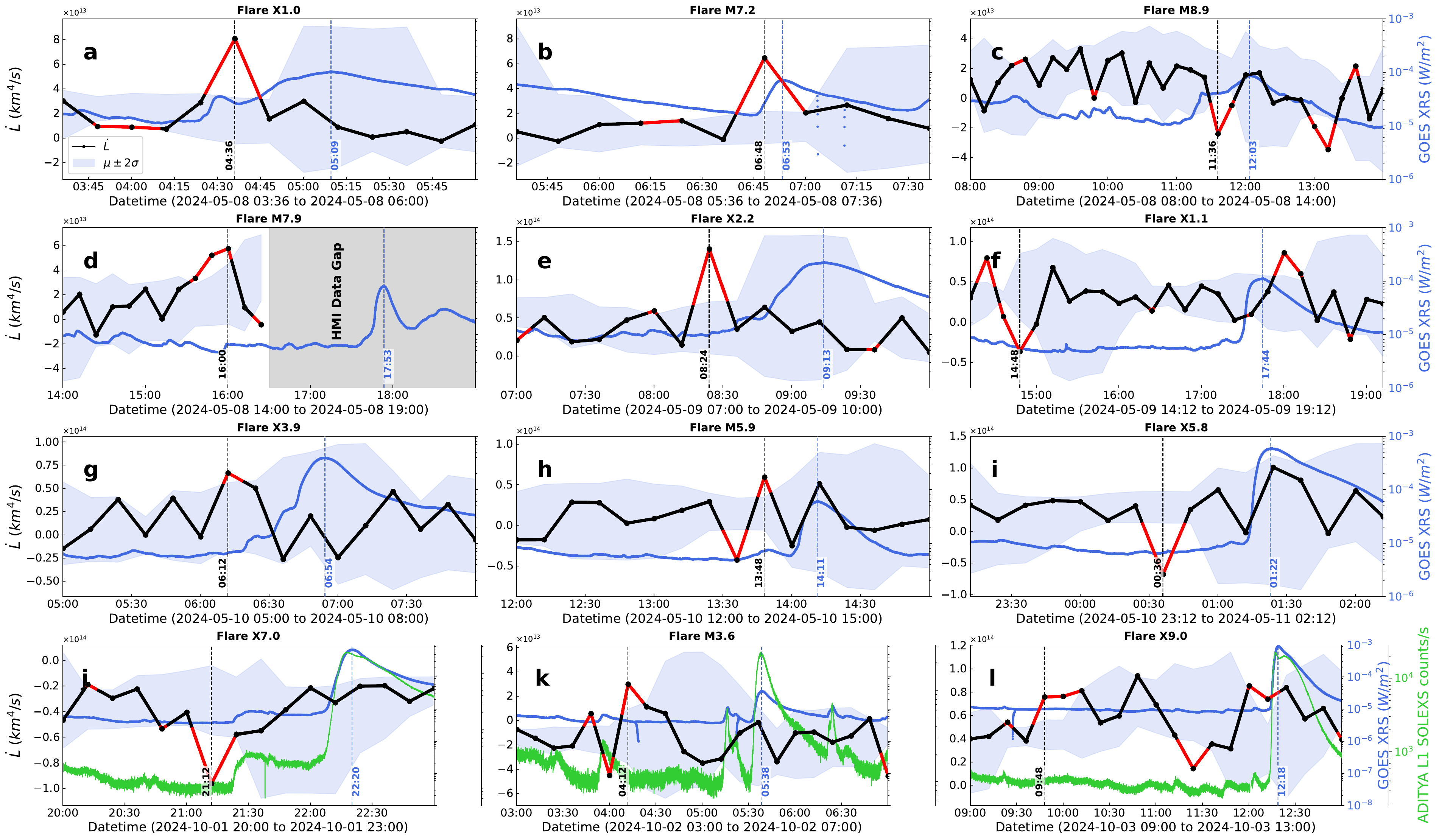}
    \caption{Zoomed-in temporal profiles of winding flux (\(\dot{L}\); a-i) for five X-class and four M-class flares from AR 13664. Panels (j, k, l): Temporal evolution for X7.1, M3.6 and X9.0 flares from AR 13842. GOES soft X-ray flux (blue) and SOLSEX X-ray counts (red; panels j, k, l) are overplotted. Flare and winding flux peaks are marked by vertical dotted blue and black lines, respectively, highlighting preflare activity.}
    \label{fig:temporal_L}
\end{figure*}
Nevertheless, the two cumulative parameters reveal distinct evolutionary behaviors that highlight their differing physical sensitivities. While helicity accumulation ($H$) exhibits a smooth, near-linear, monotonic increase that averages out localized variations, winding accumulation ($L$) displays noticeable changes in its accumulation rate. For AR\,13664, both parameters accumulate slowly during the initial phase on May 8. Following the data gap late on May 8, however, $L$ exhibits a pronounced steepening in slope, rising sharply from $\sim 0.8 \times 10^{18}$\,km$^4$ to over $3.0 \times 10^{18}$\,km$^4$ leading into the X2.2 and X1.1 flares on May 9 (Figure~\ref{fig:temporal_L_AR13664}a). This sharp slope transition in $L$ captures an episode of rapid, localized topological restructuring that the global helicity curve smooths out. Similarly, in AR\,13842, $L$ displays clear episodic changes, including a steep decrease prior to the X7.1 flare and a sharp increase leading into the X9.0 flare (Figure~\ref{fig:temporal_L_AR13842}a). As the accumulated parameters represent the cumulative background reservoir rather than high-cadence triggers, the sensitive precursors of impending flares reside in the instantaneous injection rates: the helicity flux ($\dot{H}$) and the winding flux ($\dot{L}$) (panels (b) and (c) in Figures \ref{fig:temporal_L_AR13664} and \ref{fig:temporal_L_AR13842}). While $\dot{H}$ is dominated by continuous background fluctuations, $\dot{L}$ exhibits distinct, high-amplitude excursions in either the positive or negative direction, from a few hours to minutes before flare onset.

This marked contrast between the helicity flux ($\dot{H}$) and winding flux ($\dot{L}$) stems directly from their mathematical formulations and physical weighting. Helicity flux (Equation~\ref{eq:helicity_flux}) scales quadratically with the normal magnetic field via the weighting factor $B_z(x)B_z(y)$. Consequently, $\dot{H}$ is overwhelmingly dominated by the macroscopic, large-scale motions of strong umbral flux concentrations, which effectively drowns out subtle, localized topological variations occurring near the neutral line. In contrast, the winding flux (Equation~\ref{eq:winding_flux}) replaces magnetic flux magnitudes with normalized signum functions $\sigma_z(x)\sigma_z(y)$ (Equation~\ref{eq:polarities}), isolating the purely geometric and topological rate of field-line entanglement ($d\Theta/dt$). Because pre-flare dynamics (such as localized shearing, flux cancellation, and twist buildup in the core field) involve rapid localized angular displacements along the PIL, these localized restructurings produce coherent, sharp excursions in $\dot{L}$. The polarity of these pre-flare excursions (positive or negative) is determined by the mutual sign of the interacting polarities combined with the chirality (sense of rotation and shear) of the evolving footpoints, thereby rendering $\dot{L}$ a sensitive and intensity-independent diagnostic of impending eruptive activity.

To address whether these winding flux fluctuations statistically indicate pre-flare activity rather than random noise, we identify significant excursions in $\dot{L}$. A running mean $\mu$ and standard deviation $\sigma$ are calculated, and an excursion is defined as a departure beyond the $\mu \pm 2\sigma$ envelope for AR13664 and AR13842 (green shaded region, panel c of Figures~\ref{fig:temporal_L_AR13664} \& \ref{fig:temporal_L_AR13842}). The running mean ($\mu$) is calculated from the winding-flux data over the preceding hour (corresponding to five data points), following the methodology adopted in previous studies of winding flux \citep{Mactaggart2024TheCME, Alielden2023ARTop:Photosphere, 2026asi..confO...6H, 2026cosp...46.2199M}. It should be noted that the excursion events are shown by red lines in panel (c) of Figures~\ref{fig:temporal_L_AR13664} and \ref{fig:temporal_L_AR13842}, which serve as potential pre-flare signatures. In our main manuscript, we focused on 12 representative flares from AR13664 and AR13842. We analyzed them in detail by examining their simultaneous spatio-temporal evolution to identify precursor activity and isolate the possible source region of the flares. However,  for AR 13664, we observed 36 distinct excursion events during the observational window and identified 29 flares of class M1.0 or higher. Similarly, for AR 13842, we identified 32 excursion events alongside 9 flares of class M1.0 or higher (panel c in Figures~\ref{fig:temporal_L_AR13664} and \ref{fig:temporal_L_AR13842}; Figure~\ref{fig:apjs_composite_combined}). These raw event counts provide the baseline for evaluating our false-alarm rates. However, we would like to highlight that for several flares, we have more than two excursion events, which serve as potential pre-flare signatures. For example, in AR 13664, both the M7.9 and X1.1 flares were preceded by more than two distinct excursion events. Similarly, in AR 13842, the M3.6 and X9.0 flares were each preceded by more than three excursion points (see panels d, f, k, and l for Figure~\ref{fig:temporal_L}). Furthermore, these clustered excursions appear spatially co-located around the same magnetically connected regions. This spatial and temporal clustering suggests that multiple $\dot{L}$ peaks may represent a prolonged, multi-stage build-up of winding flux before major eruptions, a physical nuance that raw statistical counting might underestimate.  \\

We tested the resulting flare-excursion (red) association against a Monte Carlo Null Hypothesis (for more details, see Appendix~A1) assuming flare occurrence times are random. The results demonstrate a statistically significant predictive relationship between the excursions and the precursors of the solar flares. We then compared the number of \textit{actual} observed flare-excursion matches (true positives) against this randomized background distribution to calculate a $p$-value. A low $p$-value ($p < 0.05$) indicates that the number of true flares following excursions within that time window is statistically significant and highly unlikely to be the result of a random distribution.
Table \ref{tab:mc_ar_combined} presents the statistical results of the Monte-Carlo test for selected prediction windows (lead time windows) for both AR 13664 and AR 13842 using a $\pm2.0\sigma$ threshold of the winding flux. For AR 13664, the predictive relationship is strongest around a 3.25-hour window ($p = 0.0032$) and for AR 13842, the relationship peaks at approximately a \textbf{1.65-hour window ($p = 0.0415$).}

\begin{figure*}
    \centering
\includegraphics[trim = 0.0cm 0.0cm 0.0cm 0.0cm, scale=0.35]{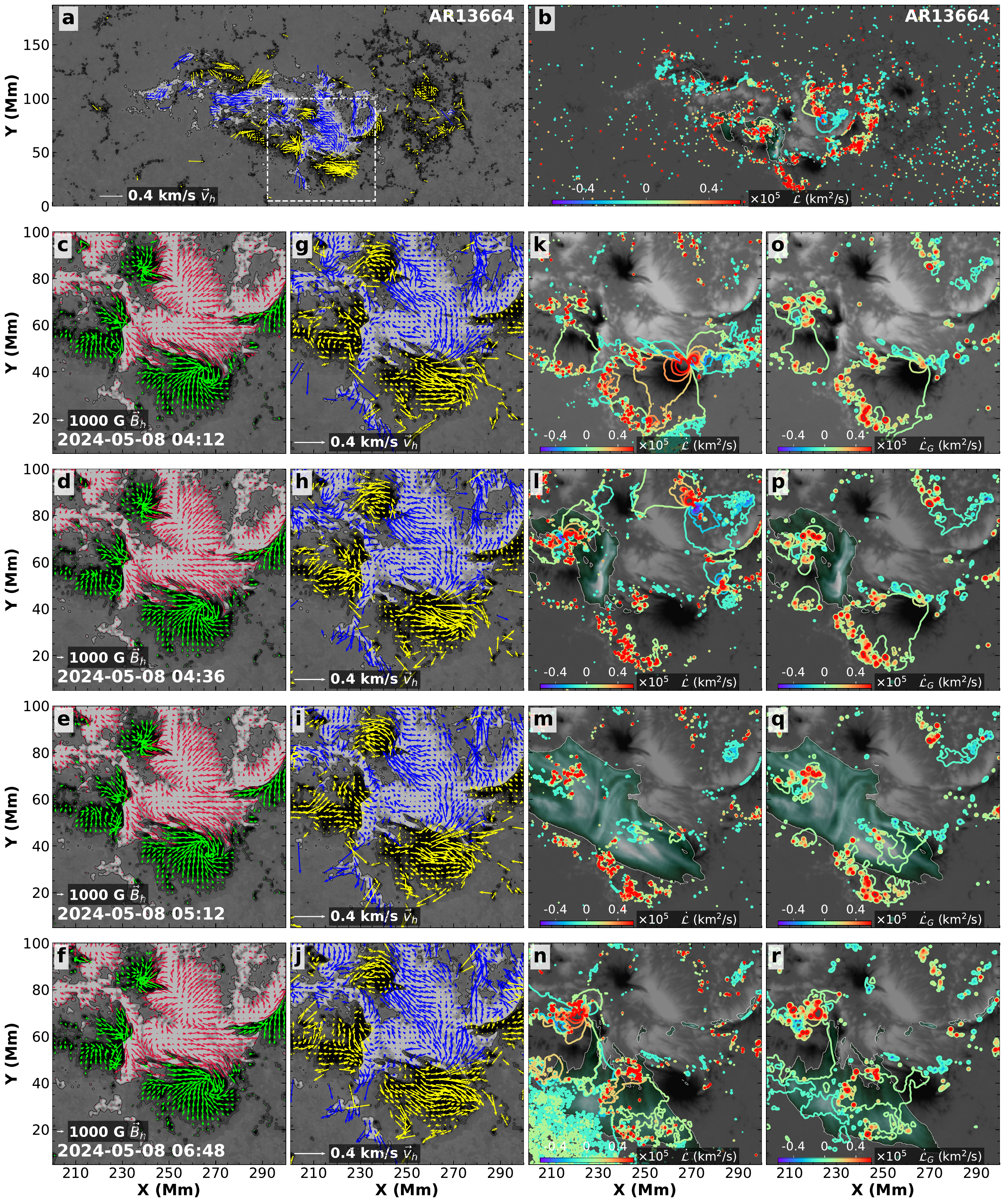}
   \caption{Spatio-temporal distribution of the magnetic and velocity fields, together with the winding flux, in AR13664 before the X1.0 and M7.2 flares. Panels (a--r) show the radial component of the magnetic field ($B_r$) derived from the HMI SHARP vector magnetograms, shown as the background, with the positive and negative magnetic polarities indicated by solid and dotted black contours, respectively. Panel (a) shows the horizontal velocity field ($\mathbf{v}_h$) overlaid on the full-field-of-view $B_r$ map. Panel (b) displays the raw winding-flux contours overlaid on the full $B_r$ map. The subsequent rows present zoomed-in views at selected pre-flare times (04:12, 04:36, 05:12, and 06:48 UT on 2024 May 8). From left to right, the columns show the horizontal magnetic-field vectors ($\mathbf{B}_h$; panels c--f), horizontal velocity vectors ($\mathbf{v}_h$; panels g--j), raw winding-flux contours (panels k--n), and the smoothed winding flux obtained using a three-dimensional Gaussian filter, overlaid on AIA 94~\AA{} intensity ($>5000$ DN s$^{-1}$; panels o--r), highlighting the spatial relationship between the winding-flux distribution and the hot coronal structures.}
    \label{fig:spatial_AR13664}
\end{figure*}
\begin{figure*}
    \centering
\includegraphics[trim = 0.0cm 0.0cm 0.0cm 0.0cm, scale=0.35]{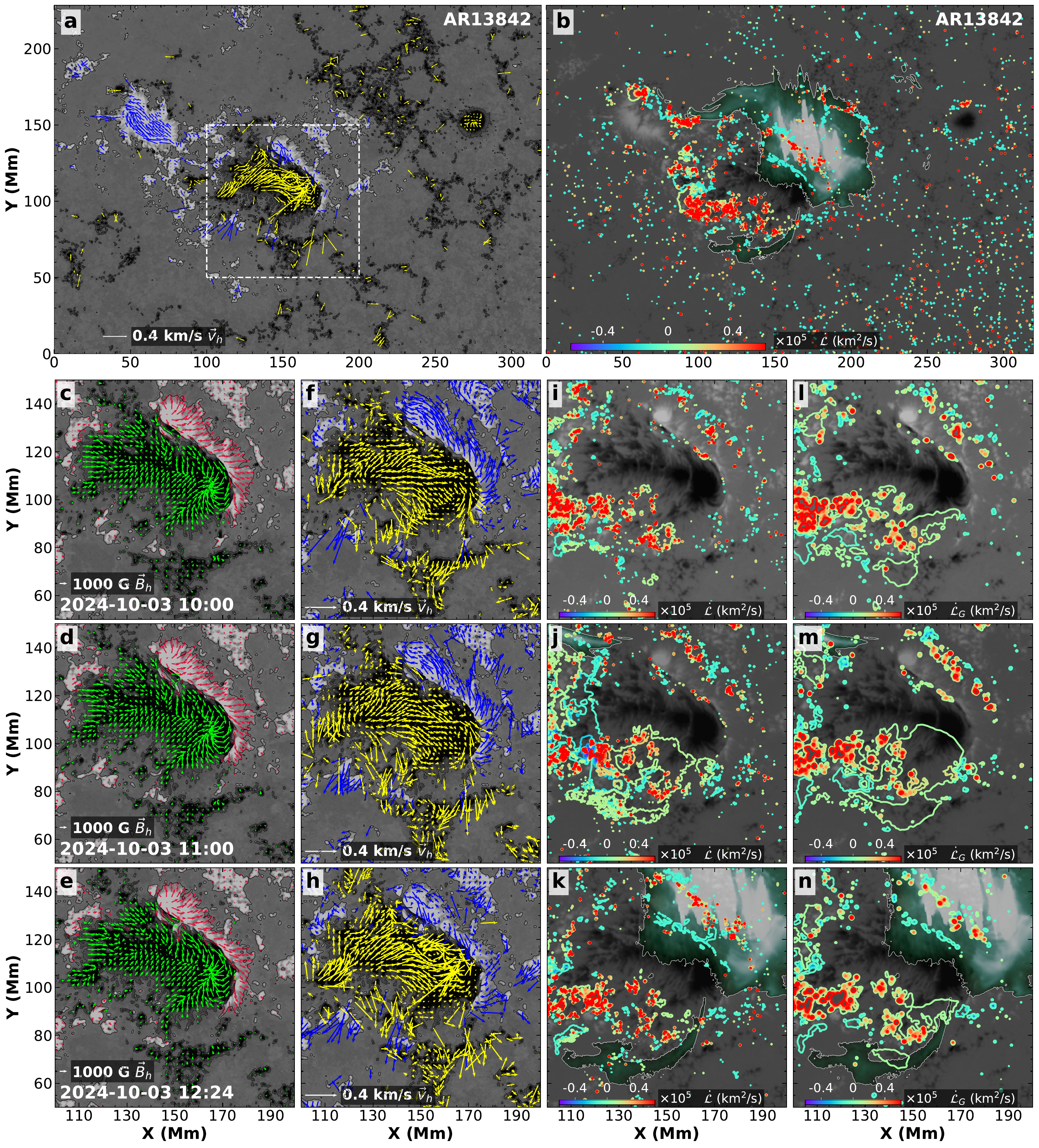}
    \caption{Spatio-temporal distribution of magnetic and velocity fields along with winding flux for the X9.0-class flare produced by AR13842. Panel (a) shows the horizontal velocity (\(\mathbf{v}_h\)) overlaid on the full field-of-view radial magnetic field (\(B_r\)). Panel (b) displays the raw winding flux contours overlaid on the full \(B_r\) map. The following rows present zoomed-in views at key pre-flare and peak times (10:00, 11:00, 12:24 UT on 2024 October 3). Columns from left to right correspond to: horizontal magnetic field vectors (\(\mathbf{B}_h\); panels c--e), horizontal velocity vectors (\(\mathbf{v}_h\); panels g--i), raw winding flux contours (panels k--m), and smoothed winding flux (3D Gaussian filtered) overlaid on AIA 94~{\AA} intensity $>$ 5000 DN s\(^{-1}\) (panels o--q), highlighting coronal connectivity.}
    \label{fig:spatial_AR13842}
\end{figure*}

Furthermore, we apply a 3D Gaussian + Normalized Convolution filter (see Appendix A2 for details) to smooth out random noise in the winding flux (blue curve in panel d of Figures \ref{fig:temporal_L_AR13664} and \ref{fig:temporal_L_AR13842}). This filtering is highly effective at reducing temporal noise and minimizing random movement in the spatial distribution of the winding flux. Such smooth variations are crucial for highlighting the magnetically dominant, highly entangled complexes (e.g., strongly sheared or twisted regions) that host intense flaring activity (last vertical column in Figures \ref{fig:spatial_AR13664} and \ref{fig:spatial_AR13842}). However, the temporal component of this 3D filter evaluates the signal over a symmetric interval $[t - \Delta t, t + \Delta t]$. Because the convolution integrates data points, temporal leakage occurs. Consequently, while this symmetric formulation is ideal for achieving retrospective visual coherence, it cannot serve directly as a strictly causal, real-time pre-flare indicator. To overcome this data-leakage in the temporal domain, we also plotted a backward-looking signal (the running mean of the preceding hour, $\mu$, shown as dashed grey lines in panel (d) of Figures~\ref{fig:temporal_L_AR13664} and \ref{fig:temporal_L_AR13842} for AR 13664 and AR 13842, respectively). Mathematically, this constrains the temporal integration strictly to the domain $t^\prime \le t$. The backward-smoothed signal reveals the distinct, localized pre-flare peaks (marked with red stars) that build up before major eruptions. Those localized peaks temporally match with the excursion region, which visually confirms the robust statistical precursor trend identified by our trailing-window analysis. The temporal evolution of the GOES X-ray flux \citep{1994SoPh..154..275G, 2005SoPh..227..231W} is overlaid on panels (b), (c), and (d), with vertical dashed lines indicating the peak times of flares $\geq$M5.0 for AR 13664 and $\geq$M1.0 for AR 13842.

Figure~\ref{fig:temporal_L} illustrates the temporal evolution of total winding flux (\(\dot{L}\)) of the studied intense flares (seven X-class flares and five M-class flares), overlaid with high-cadence GOES XRS (1~s) and Solar Low Energy X-ray Spectrometer \citep[SoLEXS;][]{2025SoPh..300...87S} onboard India's Aditya-L1 data (red curve; panels j, j, k, and l of Figure~\ref{fig:temporal_L}) for the X7.1, M3.6, and X9.0 flares in AR13842. In almost all the cases, it \(\dot{L}\) shows significant enhancements prior to the respective GOES and SoLEXS X-ray peaks. An exception occurs in the X1.1 flare, which \(\dot{L}\) peaks at 18:00~UT after the peak of the GOES X1.1 flare at 17:44 UT (panel f; Figure~\ref{fig:temporal_L}). However, the spatial distribution of the \(\dot{L}\) map (Panel a: Helicity Flux and Panel b: Winding Flux in supplementary video files S1 and S2 for AR 13664 and AR 13842, respectively) shows that this enhancement appears from a remote location, which is not magnetically connected with the possible source region of the X1.1 flare. However, we would like to highlight that an earlier enhancement appears in the spatial maps of ($\dot{\mathcal{L}}$), including a negative peak at 14:48~UT near the possible flare source region. Therefore, we suggest that this enhancement in the negative \(\dot{L}\) is a pre-indicator for the X1.1 flare. A similar negative enhancement \(\dot{L}\) appears before the X5.8 flare in AR13664 and the X7.1 flare in AR13842 (see panels f, i, and j from Figure~\ref{fig:temporal_L}). The winding flux sign strongly depends on the emergence of a positive or negative flux. These results suggest that localized winding flux enhancements, positive or negative, are a reliable preflare signature. The light-blue shaded region in each panel indicates the $\mu+2\sigma$ envelope during the precursor phase of each intense flare. The red lines corresponding to the total winding flux intersect this envelope, thereby satisfying the excursion criteria to serve as a precursor indicator (Figure~\ref{fig:temporal_L}).

\subsection{Spatial Variation of Winding Flux and Photospheric Dynamics}  
\label{sec:spatial}
The temporal evolution of the total winding flux acts as a precursor indicator for recurring flares in active regions AR13664 and AR13842 (Figure~\ref{fig:temporal_L}). To investigate the behavior of the field line winding flux during the precursor phase, we map its spatio-temporal distribution during the X1.0 and M7.2 flares from AR13664, as well as the X9.0-class flare from AR13842. Furthermore, to examine the physical context of the winding flux ($\dot{\mathcal{L}}$) during the precursor phase and peak phases, we compare its spatio-temporal evolution with those of the horizontal magnetic field ($\mathbf{B}_h$) and the horizontal velocity field ($\mathbf{v}_h$), which was derived using the DAVE4VM method (Figures~\ref{fig:spatial_AR13664} and \ref{fig:spatial_AR13842}). The raw winding flux maps (panels k--n in Figure~\ref{fig:spatial_AR13664} and panels i--k in Figure~\ref{fig:spatial_AR13842}) exhibit substantial spatial variability between successive frames. In a magnetically complex active region, different parts of the magnetic system can dominate the topological transport at different times. It should be noted that the rapid spatial displacement of localized winding signatures should not be interpreted as the rapid physical movements of a flux rope across the active region, but rather as a reflection of shifts in the locations where magnetic connectivity and topological restructuring are most intense.  Furthermore, we find that the dominant winding-flux concentrations preferentially occur in regions exhibiting enhanced horizontal flows (panels g--j in Figure~\ref{fig:spatial_AR13664} and panels f--h in Figure~\ref{fig:spatial_AR13842}) and strong transverse magnetic fields (panels c--f and c--e, respectively). Specifically, these intense winding signals align with strong shearing flows/converging motions along the primary polarity inversion lines (as outlined by the contours in panels k--n). It indicates that the winding flux is strongly associated with the evolution of magnetic connectivity, driven by photospheric flows and complex magnetic gradients, rather than reflecting the emergence of new unsigned magnetic flux. Therefore, these localized, flow-driven topological restructures appear to precursor in the active region for the subsequent flaring activity observed during the peak phase.

To confirm that the observed winding flux distribution is coherent and not dominated by pixel-scale noise, we applied a 3D Gaussian smoothing using normalized convolution to the raw $\dot{\mathcal{L}}$ (panels o--r in Figure~\ref{fig:spatial_AR13664} and panels i--n in Figure~\ref{fig:spatial_AR13842}). The smoothed $\dot{\mathcal{L}_G}$ maps reveal that the winding signal evolves on substantially longer timescales and robustly traces the coherent, large-scale structures of the pre-twisted magnetic flux. These smoothed winding enhancements spatially coincide with strong AIA 94~{\AA} brightenings ($>$~5000~DN~s$^{-1}$), confirming that winding flux effectively isolates the probable footpoints that are magnetically connected with flare-associated location. It should be noted that for the X1.0 flare in AR~13664 (peak: 05:12~UT), localized winding flux enhancements are clearly observed at 04:36~UT (consistent with \citet{Mactaggart2024TheCME}). Similarly, before the M7.2 flare, a strong enhancement in the winding flux is observed at 06:48 UT in the eastern subregion, with corresponding coronal brightenings highlighting the conjugate footpoints of the erupting flux rope. These spatial associations motivate our masked PIL-based approach discussed in Section~3.4. Check the Supplementary Video files S1 and S2 that shows the above discussed spatiotemporal variations for AR 13664 and AR13842 respectively.

\begin{figure*}[ht]
    \centering
\includegraphics[trim = 0.0cm 0.0cm 0.0cm 0.0cm, scale=0.15]{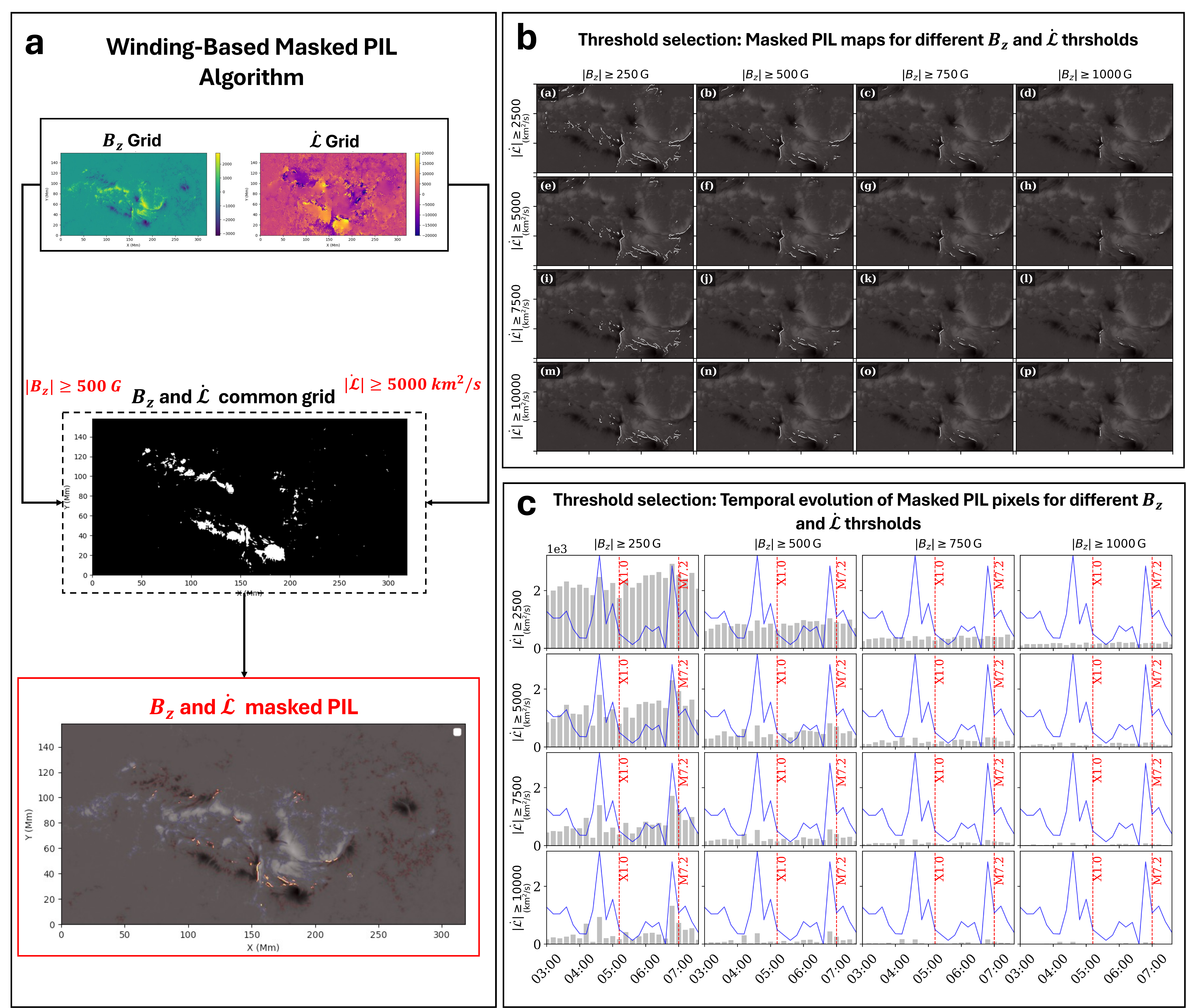}
   \caption{Panel (a) illustrates the algorithm for detecting winding-flux-based masked PILs. Panels (b) and (c) show the spatial and temporal distribution of winding-flux-based masked PILs extracted with the different combinations of \(B_z\) and \(\dot{\mathcal{L}}\) thresholds.}
    \label{fig:maskedPIL_algorithm}
\end{figure*}
\subsection{Preflare Signature near Winding-flux-based Polarity Inversion Lines} \label{masked_binned_pil}

Physics-based nonlinear force-free field (NLFFF) extrapolation \citep{Kusano2020AFlares} and the weighted horizontal magnetic field gradient (WGM) flare prediction method \citep{Korsós_2018, Korsós_2020} reveal that magnetic twists and PIL gradients—particularly near active region boundaries—provide crucial insight into the location and timing of solar flares. Recent approaches have further enhanced prediction accuracy by deriving spatial and topological features from polarity inversion line (PIL) masks applied to magnetograms and SHARP parameters \citep{2020ApJ...892..140W, Sun2021}. Also, recent studies have suggested that winding flux plays an important role as a pre-flaring indicator \citep{2022ApJ...927..156R, Mactaggart2024TheCME, Williams_2025}. Since \(\dot{\mathcal{L}}\) serves as an effective diagnostic of magnetic complexity driven by strong horizontal footpoint motions, particularly between sunspots where \(B_z\) is weak, we expect these horizontal flows to leave a clear signature in the \(\dot{\mathcal{L}}\) maps along the PILs.. To explore this, we developed a novel method for identifying PILs from SHARP magnetogram data using a combined $B_z$ and spatial winding flux $\dot{\mathcal{L}}$ masking approach. PILs are defined as regions where $B_z$ changes in polarity occur, while the $\dot{\mathcal{L}}$ mask isolates footpoint motions contributing to magnetic complexity. We further classify persistent PILs (those remaining active for $\ge 1$ hours) and strong PILs ($|\nabla B_z| > 2000$Gauss Mm$^{-1}$).

To extract the winding-flux-based masked PILs, the precalculated spatial data of \(B_z\) and \(\dot{\mathcal{L}}\) are analyzed simultaneously to determine common regions that fall within the \(B_z^{\mathrm{mask}}\) (\(B_z\) threshold) and \((\dot{\mathcal{L}}^{\mathrm{mask}} (\dot{\mathcal{L}}\ \mathrm{threshold}))\). A binary dilation mask with an extent of 2 pixels is applied to the common high-intensity regions to detect the PILs. This is done to account for small variations in the data and to ensure that adjacent regions are included, thereby allowing the algorithm to better capture regions where \(B_z\) changes occur. Following this, we extracted the PIL pixels from the masked regions for each frame. Figure~\ref{fig:maskedPIL_algorithm}(a) explains the masked PIL detection algorithm in detail. The solar magnetic field evolves dynamically and temporally at the photospheric level, as shown in the HMI magnetogram, so not all detected PILs contribute equally to flare activity. Instead, we focus on PILs that reappear consistently under the influence of higher thresholds of \(B_z\) and \(\dot{\mathcal{L}}\), providing critical insight into regions prone to flaring. Furthermore, regions with stable PILs are more likely to host sheared and twisted magnetic fields, increasing the likelihood of energy release via magnetic reconnection \citep{Liu2023ChangesFlares}.


\begin{figure*}[ht]
    \centering
\includegraphics[trim = 1.0cm 0.5cm 1.0cm 1.0cm, scale=0.85]{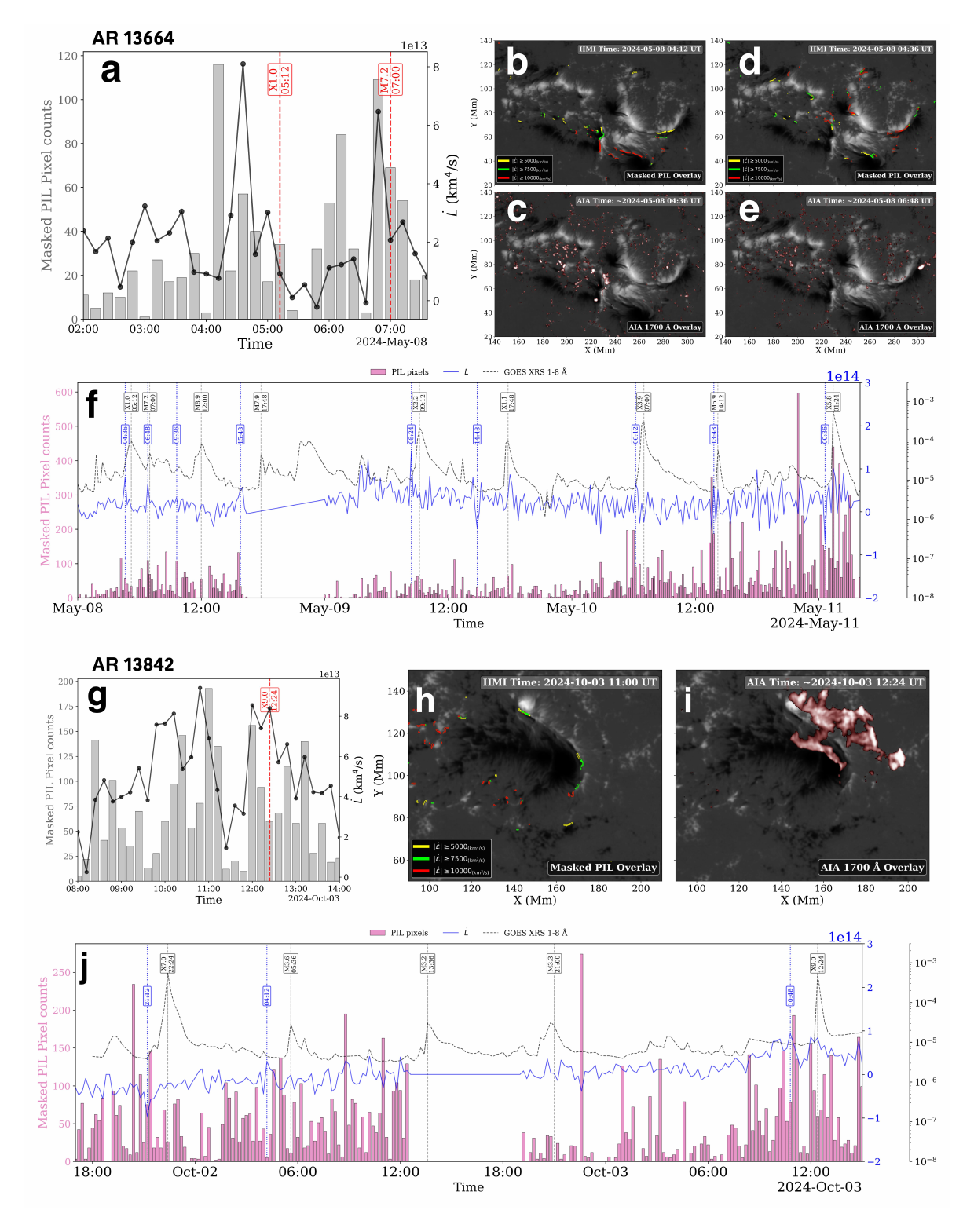}
   \caption{Panels (a, g): Simultaneous time evolution of winding flux (black line with solid circles) and winding-flux-based masked PIL pixel counts (gray bars) through the preflare-to-peak phase of X1.0 and M7.2 flares from AR 13664 and X9.0 flare from AR 13842. Red vertical \textit{dashed} lines mark flare peak times. Panels (b, d, h): Winding-based masked PIL maps showing strong PILs represented by yellow (\(|\dot{\mathcal{L}}| \geq \) 5000 km\(^2\)/s), green (\(|\dot{\mathcal{L}}| \geq \) 7500 km\(^2\)/s) and red (\(|\dot{\mathcal{L}}| \geq \)10000 km\(^2\)/s) hollow (\(|B_z| \geq \)500 G) and filled (\(|B_z| \geq \)1000 G) contours, indicating preflare activity of the flares mentioned in panels (a, g). Panels (c, e, i): Composite \(B_z\) and AIA~1700 {\AA} (red–yellow) maps, revealing flare ribbons around strong PILs during the peak time of flares mentioned in panels (a, g). Panels (f, j): Temporal evolution of winding-flux-based masked PIL pixel counts for the entire temporal domain of AR 13664 and AR 13842, indicating pre- and post-flare activity for all the flares analyzed in this study.}
    \label{fig:PIL_map}
\end{figure*}





Since the detection of PILs is strongly dependent on the vertical component of \(B_z\), they are defined as locations where \(B_z\) changes sign. To isolate \emph{strong} PILs, regions of converging opposite magnetic flux that typically host intense current density and magnetic shear \citep{Georgoulis2008, schrijver_2007}, we apply the winding flux as an additional mask to identify highly complex magnetic structures. The detection and isolation of PILs are sensitive to threshold values for both \(B_z\) and \(\dot{\mathcal{L}}\). Based on an extensive exploration of the parameter space, we adopted the appropriate thresholds of \(B_z\) and \(\dot{\mathcal{L}}\) for the isolation of the possible flux ropes, which were responsible for the launch of the intense flares, Figures~\ref{fig:maskedPIL_algorithm}(b). These optimized values enable robust detection of persistent and long-lived PIL structures. Such long-lived PILs are particularly well-suited for identifying potential flux-rope-forming regions and flare-prone regions \citep{Toriumi_2019}. The spatial maps and temporal distributions of the masked PILs with varying thresholds of \(B_z\) (250 G, 500 G, 750 G, and 1000 G) and the spatial distribution of the winding flux (\(\dot{\mathcal{L}}\)) (2500 km\(^2\)/s, 5000 km\(^2\)/s, 7500 km\(^2\)/s  and 10000 km\(^2\)/s) are shown in Figures~\ref{fig:maskedPIL_algorithm}(b and c), respectively. Lower thresholds produce noisy and fragmented PIL structures that do not reliably correspond to flare-productive regions (panel b; Figures~\ref{fig:maskedPIL_algorithm}).  By considering regions with higher winding flux, nonflare-prone regions can be effectively excluded. However, the adopted thresholds yield coherent winding flux-based PILs and enable robust extraction of preflaring PIL structures, providing a reliable basis for investigating their role in flare productivity.

Figure~\ref{fig:PIL_map} shows the spatial maps of winding-flux-based masked PILs constrained by multiple thresholds (panels b, d, h) and the temporal evolution of the corresponding winding-flux-based masked PIL pixel counts (panels a and g) for three representative flares: an X1.0- and an M7.2-class flare from AR~13664 and an X9.0-class flare from AR~13842 (for the full spatiotemporal evolution of masked PILs, see panel (a) of supplementary files S3 and S4 for AR 13664 and AR 13842, respectively). Furthermore, the full temporal evolution of the masked pixel counts is depicted in Figures~\ref{fig:PIL_map} (f and j) respectively for AR13664 and AR13842. To highlight the post-flare response, we overlay AIA 1700~\AA\ contours on the corresponding HMI maps (pink contours; panels~c, e, and h). Masked PILs represented by yellow (\(| \dot{\mathcal{L}} | \geq 5000\)~km\(^2\)/s), green (\(| \dot{\mathcal{L}} | \geq 7500\)~km\(^2\)/s), and red (\(| \dot{\mathcal{L}} | \geq 10000\)~km\(^2\)/s) hollow (\(|B_z| \geq 500\)~G) and filled (\(|B_z| \geq 1000\)~G) contours are shown in panels~(b), (d), and (h) and are found to intensify around the post-flare loops of AIA 1700~\AA\ shown as pink contours overlaid on the corresponding HMI maps in panels~(c), (e), and (i) of Figure~\ref{fig:PIL_map}.

The masked PIL pixel counts exhibit a significant increase that is positively correlated with the enhancement of the winding flux (panels (a) and (d); Figure~\ref{fig:PIL_map}). A similar increase was observed for all other flares analyzed (X2.2, M8.9, M7.9, X1.1, X3.9, M5.9, X5.8, X7.1, M3.6, M3.2 and M3.3; see panels f and j in Figure~\ref{fig:PIL_map}). In practice, both the winding flux and the masked PIL pixel counts are enhanced nearly simultaneously during the preflare phase of the flares, suggesting that these parameters may provide valuable indicators for flare forecasting. These results consistently demonstrate that the masked PIL structures intensify either prior to or together with the winding flux well before the onset of a flare. Moreover, the spatial overlap of flare ribbons (pink contours) with masked PILs corresponding to high \(B_z\) and \(\dot{L}\) thresholds (yellow, green, and red) hollow highlights regions of intense magnetic shear and flux emergence, which may be associated with the possible flux ropes. Therefore, the combined use of winding flux and derived masked PIL provides a reliable framework for identifying flare-prone regions in ARs and understanding the buildup of shear and twist that leads to the launch of intense solar flares.




\subsection{Gradient of Strong PIL}

To further refine the spatial pinpointing of potential flare sites in the ARs, we computed the gradient of the masked PIL, defined as,
\begin{equation}
\nabla\mathrm{(PIL)}(x,y) \;=\;
\sqrt{\left(\frac{\partial B_z}{\partial x}\right)^2 + \left(\frac{\partial B_z}{\partial y}\right)^2},
\label{eq:gradpil}
\end{equation}
evaluated only along the winding-flux-based masked PIL segments. The physical basis for this approach lies in the fact that strong magnetic gradients at the PIL correspond to nearby steeply opposing magnetic flux systems, which are prone to magnetic reconnection once the coronal field becomes stressed. Concentrated \(\nabla\mathrm{PIL}\) patches indicate regions of large free magnetic energy accumulations, making them natural candidates for flare initiation \citep{falconer_2008, Schrijver_2008, Korsós_2018, Korsós_2020}.

\begin{figure*}[p]
    \centering
\includegraphics[trim = 0.0cm 0.0cm 0.0cm 0.0cm, scale=0.55]{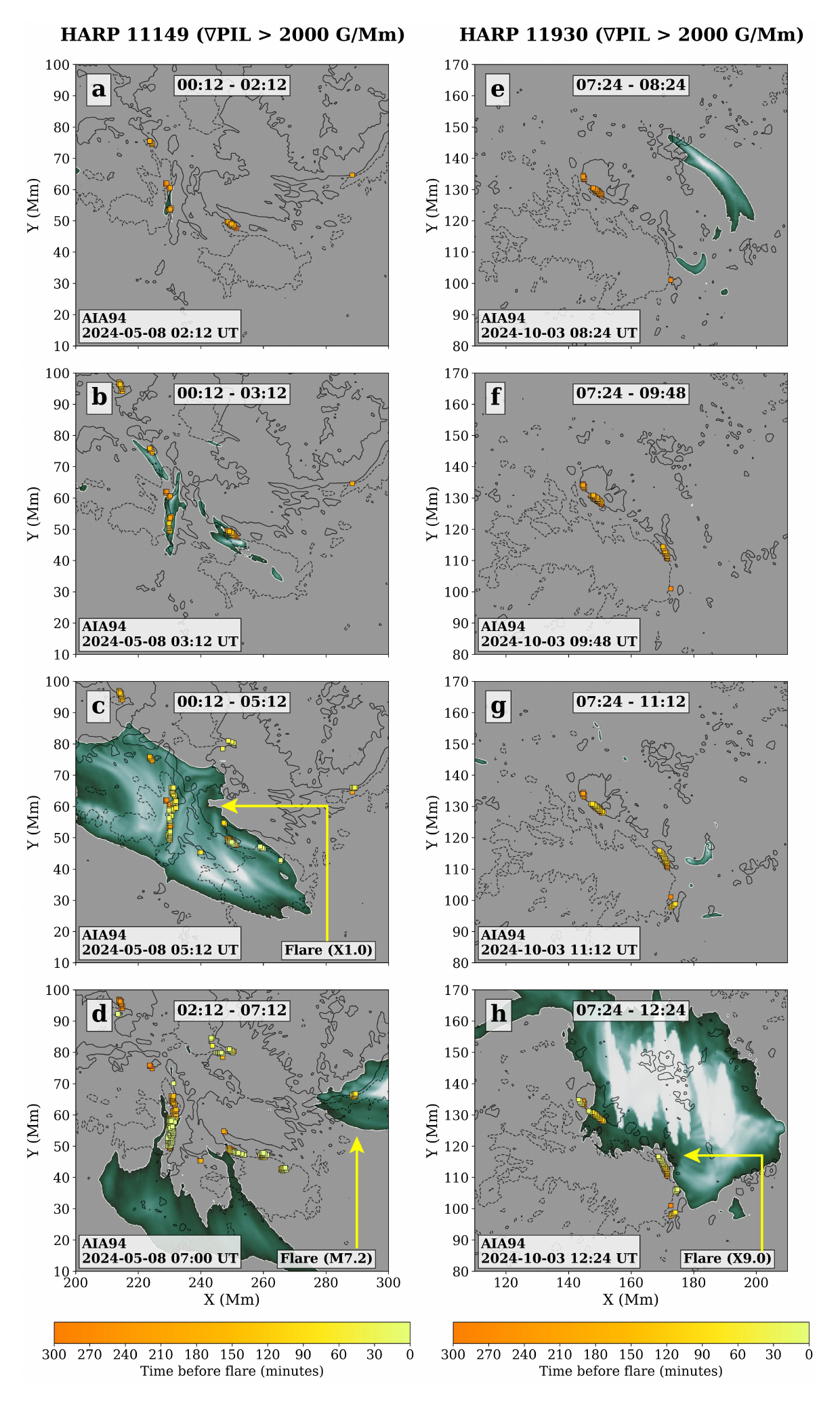}
   \caption{Spatiotemporal evolution of strong \(\nabla(PIL)\) pixel clustering before flare onset, derived from winding-based masked PIL maps for X1.0, M7.2 (AR13664, left) and X9.0 (AR13842, right). Colored tiles indicate locations where winding-based masked \(\nabla(PIL)\) exceeds 2000 G Mm$^{-1}$. Coronal connectivity is traced by overplotted AIA 94 {\AA} intensity (black loops). Clustering and temporal growth of high-gradient PIL segments highlight pre-flare magnetic geometry and localization of intense activity.}
    \label{fig:grad_PIL}
\end{figure*}

Previous studies suggest that strong magnetic field gradients across PILs are positively correlated with the possible locations of flare trigger sites \citep{2008ASPC..383..429W, 2024ApJ...964..159L}. To isolate these regions, we apply a threshold to the PIL gradient, thereby identifying locations of strong PILs,
\begin{equation}
\nabla\mathrm{(PIL)}(x,y) \;=\;
\begin{cases} 
1 & \text{if } \nabla\mathrm{(PIL)}(x,y) \geq 2000, \\
0 & \text{otherwise.}
\end{cases}    
\end{equation}
Figure~\ref{fig:grad_PIL} shows the spatio-temporal distribution of strong PIL ($\geq$2000 G/Mm) clustring for X1.0 and M7.2 class flare from AR13664 and X9.0 class flare from AR13842. We tracked the number of high \(\nabla\mathrm{PIL}\) pixels (\(>2000~\mathrm{G\, Mm^{-1}}\)) for AR13664 and AR13842 in the whole temporal domain covering all the major flares (see panels b and c from the supplementary video files S3 and S4). The method consistently detected emerging high-gradient PIL pixel clusters 3-4 hours before flare onset, with occasional early scattered signals up to 5 hours in advance (\emph{left} panels a-c for the X1.0-class flare, \emph{left} panel d for the M7.2 flare from AR13664, and {\emph{ right } panels e-h} for the X9.0-class flare from AR13842). The overlying AIA 94 {\AA} coronal loops intensified around the surrounding locations, where the strong gradient of PIL clusters is developing. The clustering intensified and became spatially confined within the flaring region during the final hour, suggesting possible trigger sites for the flares.

\begin{figure*}[p]
    \centering
\includegraphics[trim = 0.0cm 0.0cm 0.0cm 0.0cm, scale=0.56]{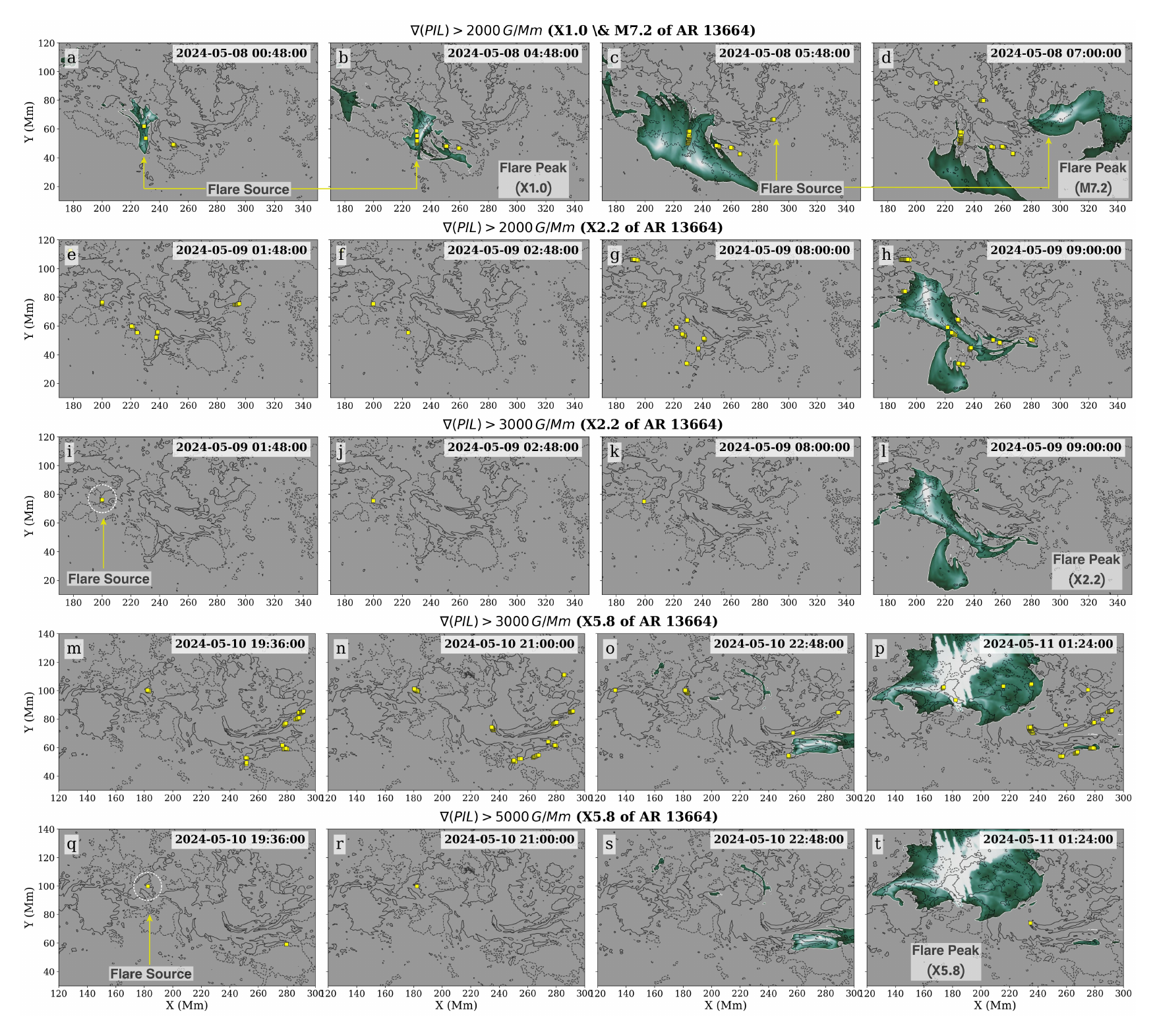}
   \caption{Spatiotemporal evolution of strong \(\nabla(PIL)\) pixels before flare onset, derived from winding-based masked PIL data for three major flares in AR13664, isolates potential triggering sites during recurrent activity. Panels (a–d) show \(\nabla(PIL)\) evolution for X1.0 and M7.2 using a threshold of 2000  G Mm$^{-1}$. Panels (e–h) and (i–l) present X2.2 evolution at 2000  G Mm$^{-1}$ and 2500  G Mm$^{-1}$, respectively, while panels (m–p) and (q–t) show X5.8 at 2000 G Mm$^{-1}$ and 5000  G Mm$^{-1}$. Varying the threshold allows identification of possible triggering sites (white circles) associated with the recurring flares.}
    \label{fig:grad_high_PIL}
\end{figure*}

\subsection{Detection and Characterization of Recurring Flare Trigger Sites Using Winding-Based PILs}
AR 13664 produced five X-class flares and several M-class flares within the span of four days. Such recurring, high-energy flaring activity poses a significant challenge for flare-forecasting models. With each flare launch, the magnetic field undergoes substantial reconfiguration, altering the pre-flare signatures on which these models depend. Therefore, the predictability of most flare-forecast methods is dramatically reduced in the presence of recurring flaring activity. For example, statistical and machine-learning-based models as well as physics-based models such as \citet{2003ApJ...595.1277L, 2019ApJS..243...36L}, \citet{Bobra_2015}, \citet {2015ApJ...802L..21K, Korsós_2018}, \citet{2017ApJ...835..156N}, and \citet{Kusano2020AFlares}, which depend on magnetic features extracted from single-time snapshots (e.g., SHARP parameters, magnetic gradient metrics), often underperform when the coronal magnetic configuration evolves rapidly after successive flares. This effect may occur because magnetic shear and twist near the PILs can be rapidly redistributed after each flare, altering the local magnetic environment and undermining forecast reliability. These limitations underscore the need for time-dependent flare forecasting methods that explicitly consider magnetic energy recovery and restructuring in active regions following major flares.

To address the difficulty in identifying the possible source regions of recurring flaring sites, we apply a threshold criterion based on the high-gradient segments of the PILs. Figure~\ref{fig:grad_high_PIL} shows the three representative X-class flares produced by AR13664. Yellow circles indicate strong-gradient PIL pixels. The top row (panels a-d) shows the spatial distribution of high-gradient PIL (\(\nabla\mathrm{PIL} > 2000 G Mm^{-1}\)) pixels during the X1.0- and M7.2-class flares at different times. As discussed above, the high-gradient PIL pixels clustered near the possible flare-triggering site (see also Figure~\ref{fig:grad_PIL} and supplementary video files S3 and S4). The second row (panels e–h) displays the spatio-temporal distribution of PIL gradients exceeding \(2000~\mathrm{G\, Mm^{-1}}\) during the X2.2-class flare. In this case, the strong-gradient PIL pixels are more scattered and do not cluster around the triggering site. To better isolate the probable flare site (white circle), we increase the threshold to \(>2500~\mathrm{G\, Mm^{-1}}\) (panels i–l), which successfully localizes the strong-gradient PIL pixels around the possible triggering site for X2.2 class flare.  A similar scenario is observed for the X5.8-class flare, where applying a higher threshold of \(>4000~\mathrm{G\, Mm^{-1}}\) isolates the strong-gradient PIL pixels around the possible flare initiation site (white circle) (panels m-p shows distribution of \(\nabla\mathrm{PIL} > 2000 G Mm^{-1}\) and panels q-t shows the same for \(\nabla\mathrm{PIL} > 5000 G Mm^{-1}\)). Therefore, in conclusion, by controlling the threshold value of PIL gradient (\(\nabla\mathrm{PIL}\)) derived from the masking of winding flux, we are able to isolate the possible trigger sites of recurring flaring activities.

\section{Discussions}

\begin{table*}
\begin{center}
\begin{minipage}{\textwidth}
\caption{Temporal Characteristics of X-Class Flares in AR13664 and AR13842: Summary of flare peak times (\(T_f\)), winding flux peak time  (\(T_{\dot{L}}\)), winding-flux-based masked PIL that captured pre-flare signature (\(T_{WPIL}\)), time difference between flare peak time and winding flux peak time (\(T_f - T_{\dot{L}}\)), time differences between flare peak time and winding-flux-based masked PIL peak time (\(T_f - T_{WPIL}\)) $\geq$ (\(T_{\dot{L}}\)) and the time differences between flare peak time and the time of accumulated clusters of high gradient of PIL \(T_f - T_{CGPIL}\).}\label{tab5}
\scriptsize
\begin{tabular*}{\textwidth}{@{\extracolsep{\fill}}lccccccc@{\extracolsep{\fill}}}
\hline

NOAA AR & Flare & \(T_f\)  & \(T_L\) & \(T_{WPIL}\)  & \(T_f - T_L\) & \(T_f - T_{WPIL}\) & \(T_f - T_{CGPIL}\)\\
\hline
\hline
AR 13664	&	X1.0	&	2024-05-08 05:12	&	04:12	&	04:12	&	60 min & 60 min & \(\approx 5 - 7\) Hrs\\
    &   M7.2        & 2024-05-08 06:53         &     06:48  &   06:48     &  8 min  & 8 min & \(\approx 4 - 6\) Hrs\\
    &   M8.9        & 2024-05-08 12:03         &     09:36  &   09:36     &  156 min  &  256 min & \(\approx 2 - 3\) Hrs\\
    &   M7.9        & 2024-05-08 17:53         &     15:48  &   15:36     &  120 min  &  132 min & \(\approx 2 - 3\) Hrs\\
	&	X2.2    	& 2024-05-09 09:12	     &	08:24	&	08:24	&	48 min & 48 min & \(\approx 4 - 6\) Hrs\\
	&	X1.1	&	2024-05-09 17:48   &	14:48	&	14:48	&	180 min & 180 min &  \(\approx 4 - 6\) Hrs\\
	&	X3.9	&	2024-05-10 07:00 &	06:12	&	06:00	&	48 min & 60 min & \(\approx 5 - 7\) Hrs \\
    &   M5.9        & 2024-05-10 14:11         &     13:48  &   13:36    &  24 min  &  36 min & \(\approx 2 - 3\) Hrs\\
	&	X5.8	&	  2024-05-11 01:24     &	00:36	&	00:36	&	48 min & 48 min  & \(\approx 6 - 8\) Hrs\\
AR 13842	&	X7.1	&	  2024-10-01 22:24     &	 21:12	&	20:24	&	72 min & 120 min & \(\approx 6 - 8\) Hrs\\
	&	M3.6	&	  2024-10-02 05:38     &	 04:12	&	03:36	&	84 min & 108 min & \(\approx 3 - 4\) Hrs\\
	&	X9.0	&	  2024-10-03 12:24     &	 10:48	&	10:24	&	96 min & 120 min & \(\approx 7 - 9\) Hrs \\
\hline

\end{tabular*}

\end{minipage}
\end{center}
\end{table*}

The main objective of this work is to investigate the spatiotemporal evolution of key magnetic topological parameters during the preflaring phase of intense solar flares. Our spatio-temporal analysis reveals that the winding flux shows a significant enhancement during the preflare phase of the flares, and it is spatially positively correlated with pre-existing flux ropes. It highlights the use of winding flux as a preflare indicator for flare forecasts. Earlier studies have demonstrated that intense winding flux reliably marks highly twisted, sheared magnetic flux ropes near PILs, which are key sites for the accumulation and eventual release of large amounts of magnetic energy in major solar flares \citep{2022ApJ...927..156R, Williams_2025}.

On the other hand, we did not find any significant pre-flaring enhancement of helicity flux ($\dot{H}$) in the spatiotemporal analysis of both the ARs. However, various studies show that the significance of helicity flux lies in its role in interpreting free magnetic energy accumulation and the subsequent launch of intense flares and CMEs \citep{Park2010a, 2010ApJ...718...43P, 2012SoPh..277..165K, Liu2023ChangesFlares, 2025arXiv250919864M, 2024A&A...686A.148S}. In short, magnetic helicity, which quantifies the global storage, linkage, and topology of magnetic field lines in active regions, is strongly weighted by field strength and serves as a key measure of the buildup of free magnetic energy. However, the winding flux ($\dot{L}$), a normalized form of helicity, captures the purely topological component of field-line connectivity, effectively separating geometric complexity from flux strength, and provides a localized, intensity-independent diagnostic of the twisted flux rope that characterizes the entanglement of field lines. Furthermore, winding flux effectively traces the emergence of twisted flux ropes, revealing topological complexity even before their complete formation \citep{Prior2020MagneticFor, MacTaggart2021HelicityEmergence, 2019JPlPh..85b7701P, MacTaggart2021DirectRegions}. This enhanced topological sensitivity makes winding flux (\(\dot{\mathcal{L}}\)) particularly valuable for identifying pre-flare signatures and tracking the evolving magnetic complexity of active regions.\\

\citep{Sun2021} found that clustering of the radial field ($B_r$) around PILs better captures pre-flare evolution than other SHARP parameters by masking the magnetogram. Statistical analyses of 40 active regions and more than 200 flares show that relative helicity near PILs significantly decreases after intense flares \citep{2024A&A...690A.181M, 2025arXiv250919864M}. Existing flare forecast models use feature extraction methods, such as unsigned flux, current helicity, and PIL structure, to capture active region dynamics and use them as a preflare activity for flare forecast \citep{Ahmadzadeh2021HowEvents, Bhattacharjee2020SupervisedMagnetograms, 2021ApJ...910....8Y, Sun2022SolarNetworks, Yi2023ApplicationForecasting}. The Keys flare forecast models (e.g., MAG4, WGM, ML classifiers, and the $\kappa$-method) use ML-based (e.g., logistic regression, CNNs) and physics-based approaches that use LOS and SHARP magnetograms, GOES flux, and AIA data to predict pre-flare activity through analysis of magnetic energy, topology, gradients, and MHD instability, \citep{2014SpWea..12..306F, Bobra_2015, 2015ApJ...802L..21K, Korsos_2016, 2017ApJ...835..156N, Kusano2020AFlares, 2021EP&S...73...64N}. While effective, these methods are computationally demanding, data-intensive, and sometimes lack physical interpretability.\\

In this paper, our winding-flux-based PIL approach offers a novel method for isolating regions of complex magnetic topology and emerging twisted flux ropes. The winding-based masked PILs peak during the preflare phase and are spatially clustered around regions of strong shear and flux accumulation, suggesting possible triggering sites in the ARs. Controlling the PIL gradient threshold based on PILs extracted from winding flux masking improves the localization of flare initiation sites in both confined and recurring activity. Incorporating topological parameters from observations with machine learning enhances the predictability and physical interpretability of solar flare forecasts, even in recurring regions where many existing models lose accuracy. In our ongoing and future work, we are conducting statistical analyses of various flare classes and AR types, with a focus on the predictive role of magnetic topology-based PIL extraction, particularly winding flux and helicity flux, as pre-flare indicators. We are analyzing confined and recurring flaring ARs to isolate magnetic structures, identify flare sites, and estimate onset times by varying the PIL gradient threshold based on the winding flux, as discussed here for intense X-class flares. This methodology offers a physically grounded feature for machine-learning-based flare prediction, thereby enhancing forecasts of flare locations and recurrence. In the future, our physics-based extracted topological parameters, combined with machine learning and additional SHARP and AIA parameters, will improve the predictability and interpretability of flare forecasts. Extracting topological features bridges the gap between data-driven learning and physical insight, thereby enhancing our understanding of preflare and eruption sites and their physical interpretability.

\section{Acknowledgment}
The authors sincerely thank the anonymous reviewer for their constructive comments and valuable suggestions, which significantly improved the quality and clarity of this manuscript. P.H., an adjunct faculty member in the Department of Physics and Astronomy at the University of Nebraska–Lincoln, sincerely thanks the institution for generous access to its facilities, including library and research resources, which greatly supported this work. S.K.M. acknowledges Kyoto University, Japan, for providing him with an institute fellowship and computational facilities. A.K.S.  acknowledges the support of ANRF, India, for providing a research grant to carry out the research (Ref:- ANRF/ARG/2025/008017/PS). Data courtesy of NASA/SDO and the HMI and AIA science team. We also acknowledge the use of the ARTop code \citep{Alielden2023ARTop:Photosphere} to extract the topological parameters of the two ARs. This work was completed utilizing the Holland Computing Center of the University of Nebraska, which receives support from the UNL Office of Research and Innovation and the Nebraska Research Initiative.

\begin{contribution}
P.~Halder contributed to the data analysis, code development, investigation, error analysis, mitigation, manuscript writing and science. S.K.~Mishra contributed to the interpretation of the results and the writing of the manuscript. A.K. Srivastava and A. Asai contributed to manuscript reading, editing, and science. P. A. Dowben contributed to the technical discussions and the manuscript review.
\end{contribution}

\section{Data Availability}
All the related supplementary animation files mentioned in the manuscript are available on Zenodo through this link (\url{https://doi.org/10.5281/zenodo.17941100}) \citep{halder_2026_17941100}.

\appendix
\renewcommand{\thesection}{A\arabic{section}}
\setcounter{figure}{0}
\setcounter{table}{0}
\renewcommand{\thefigure}{\thesection.\arabic{figure}}
\renewcommand{\thetable}{\thesection.\arabic{table}}
\renewcommand{\theHfigure}{\thesection.\arabic{figure}}
\renewcommand{\theHtable}{\thesection.\arabic{table}}
\makeatletter
\def\fnum@figure{{\bf\figurename~\thefigure}}
\def\fnum@table{{\bf\tablename~\thetable}}
\makeatother
\section{Monte-Carlo Null Hypothesis Test}
\label{sec:appendix_mc}

\begin{figure*}[ht]
    \centering
    \includegraphics[trim = 0.0cm 0.8cm 0.0cm 0.0cm, scale=0.8]{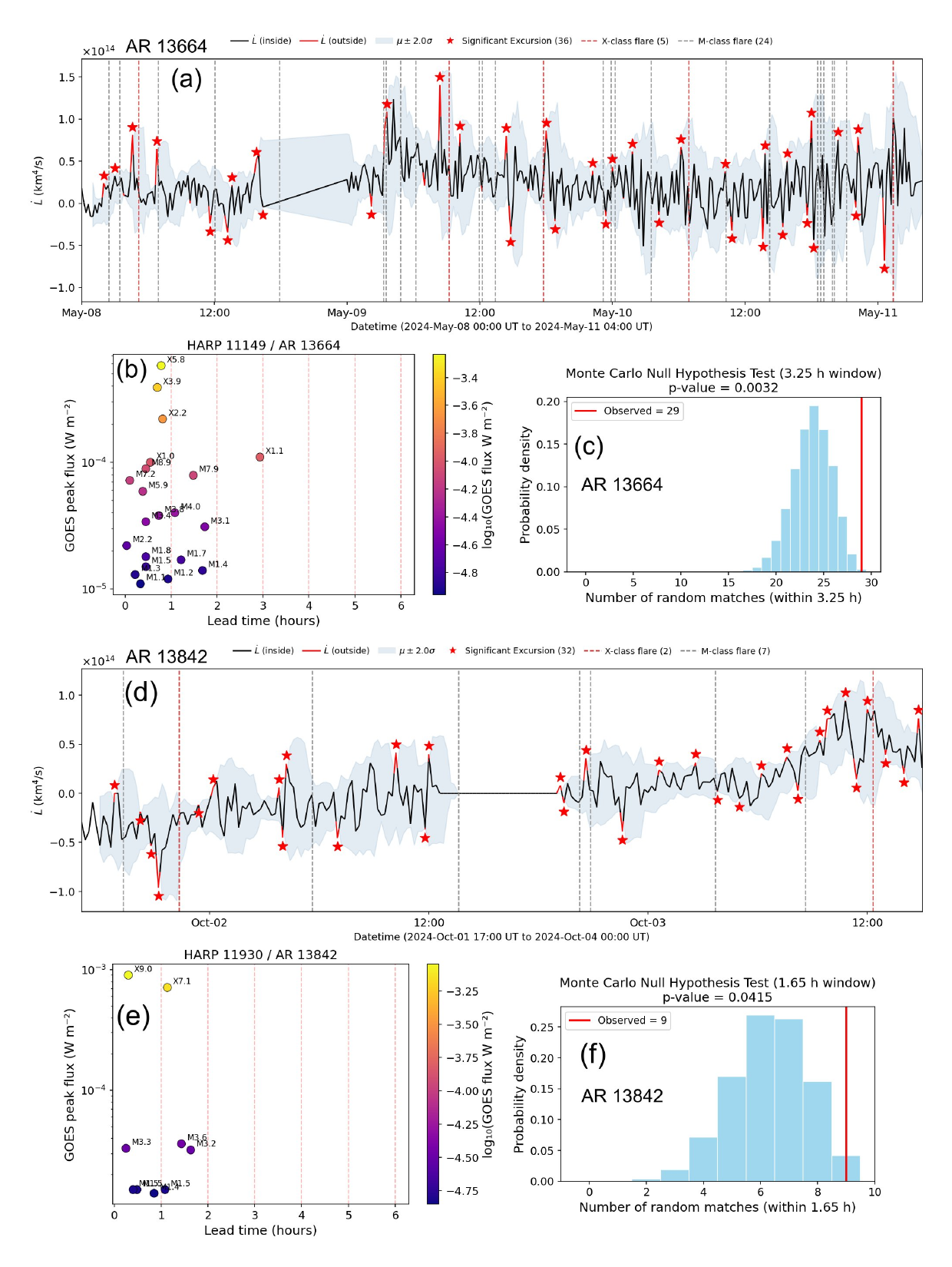}
    \caption{Monte-Carlo Null Hypothesis Test results and flare associations. (a) Temporal evolution of the winding flux for AR 13664, with identified excursions (red stars) exceeding the $\pm2.0\sigma$ threshold (blue shaded region) and observed flares $\ge$M1.0 (vertical dashed lines). (b) GOES peak flux versus lead time for flares in AR 13664. (c) Monte-Carlo simulation histogram at the peak significance window (3.25 hours) for AR 13664, comparing the number of observed flare associations (red line) to the randomized background distribution (blue). Panels (d), (e), and (f) show the corresponding temporal evolution, lead times, and Monte-Carlo test results (1.65 hours window) for AR 13842, using the same $\pm2.0\sigma$ excursion threshold.}
    \label{fig:apjs_composite_combined}
\end{figure*}

To validate the predictive relationship between winding flux excursions and solar flares, Figure \ref{fig:apjs_composite_combined} illustrates our excursion detection and statistical testing framework. 
First, we establish a causal running mean $\mu$ and a dynamic threshold envelope ($\mu \pm 2.0\sigma$) over the temporal evolution of the winding flux (panels a and d). Significant excursions are registered when the raw flux exceeds this threshold, with local extrema marked as red stars. We then map the lead times measured from these excursion peaks to the onset of subsequent intense flares ($\ge$M1.0) against the GOES peak flux (panels b and e). This visualizes the precursor temporal windows leading up to various flare magnitudes.
To evaluate whether these associations occur by random chance, we employed a Monte-Carlo null hypothesis test. Detected winding flux excursions were fixed in time while flare onset times were randomly shuffled 10,000 times (deterministically seeded for reproducibility) across the observation window. The resulting probability distributions of random flare-excursion matches are shown in panels (c) and (f) at the peak prediction windows (3.25\,h for AR 13664; 1.65\,h for AR 13842). 

\begin{table}[ht]
\centering
\caption{Monte-Carlo Null Hypothesis Test Results for AR 13664 and AR 13842 ($\pm2.0\sigma$ threshold)}
\label{tab:mc_ar_combined}
\begin{tabular}{cccc}
\hline
\textbf{Window (hours)} & \textbf{Observed} & \textbf{Random Mean} & \textbf{$p$-value} \\
\hline
\multicolumn{4}{c}{\textbf{AR 13664}} \\
\hline
1.00 & 17 & 11.39 & 0.0261 \\
2.00 & 26 & 19.17 & 0.0038 \\
3.00 & 28 & 23.04 & 0.0102 \\
3.25 & 29 & 23.78 & \textbf{0.0032} \\
4.00 & 29 & 25.30 & 0.0191 \\
\hline
\multicolumn{4}{c}{\textbf{AR 13842}} \\
\hline
1.00 & 5 & 4.85 & 0.5928 \\
1.65 & 9 & 6.35 & \textbf{0.0415} \\
2.00 & 9 & 6.90 & 0.0912 \\
3.00 & 9 & 7.67 & 0.2406 \\
4.00 & 9 & 8.02 & 0.3536 \\
\hline
\end{tabular}
\end{table}

A $p$-value is calculated as the fraction of random iterations yielding matches greater than or equal to the observed count. Table \ref{tab:mc_ar_combined} presents these statistical results. A low $p$-value ($p < 0.05$) confirms the predictive relationship is highly significant and non-random, peaking at $p=0.0032$ (3.25\,h window) for AR 13664 and $p=0.0415$ (1.65\,h window) for AR 13842.

\setcounter{figure}{0}
\setcounter{table}{0}
\section{3D Gaussian Blur + Normalized Convolution}
\label{sec:appendix_a2}

\begin{figure*}[ht]
    \centering
    \includegraphics[width=\textwidth]{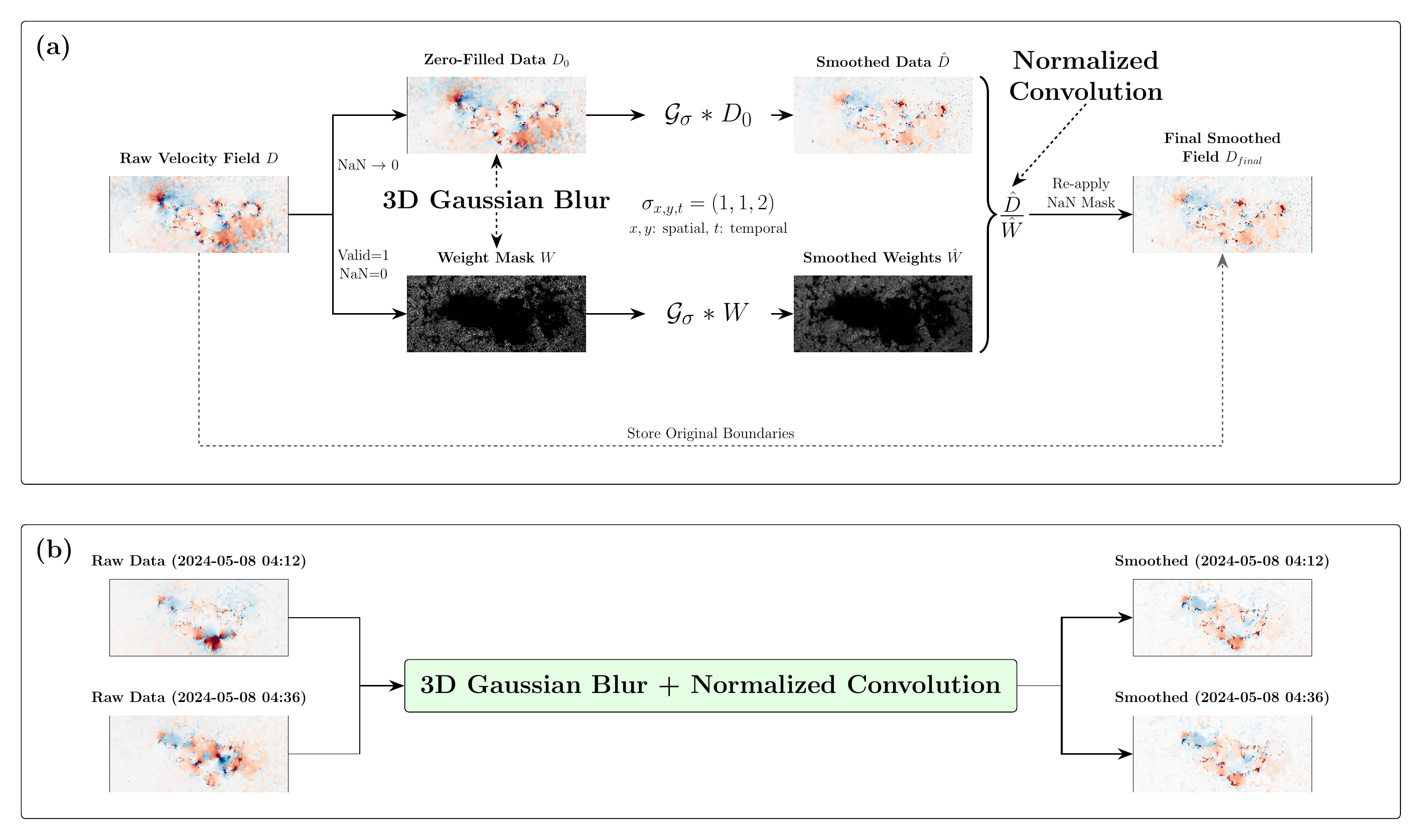}
    \caption{Methodological flowchart illustrating the 3D Gaussian blur and normalised convolution applied to the temporal active region data. Panel (a) describes the mathematical operations on the data ($\hat{D}$) and weight ($\hat{W}$) volumes. Panel (b) provides visual examples of the raw versus smoothed maps, demonstrating the noise reduction and signal preservation.}
    \label{fig:appendix_a2_smoothing}
\end{figure*}

To mitigate the effects of high-frequency noise and handle missing data points (NaNs) in the continuous observational cadence, we applied a combination of 3D Gaussian Blur and Normalised Convolution to the time series of spatial maps (see Figure \ref{fig:appendix_a2_smoothing}). First, a binary weight mask $\hat{W}$ is generated, where valid data pixels are assigned a value of 1 and NaN pixels are assigned a value of 0. The raw data maps, $\hat{D}$, are prepared by setting NaN values to 0. We then apply a 3D Gaussian Blur with a standard deviation $\sigma_{x,y,t}=(1,1,2)$ simultaneously to both the data volume $\hat{D}$ and the weight volume $\hat{W}$. The smoothing over the temporal dimension ($t$) helps to enforce consistency across consecutive frames, while the spatial smoothing ($x,y$) reduces local pixel-level noise. 

To restore the true signal intensity, especially near the boundaries of missing data regions, we compute the Normalized Convolution as the element-wise division of the blurred data by the blurred weights:$\hat{D}_{smooth} = \frac{(\hat{D} * G)}{(\hat{W} * G)}$. Finally, the original NaN mask is reapplied to the smoothed data to ensure that the physical boundaries of the active regions are strictly preserved, preventing artificial inflation of the active region area.

\bibliography{references}{}
\bibliographystyle{aasjournal}




\end{document}